\documentclass{article} 
\usepackage{emulateapj}
\usepackage{psfig}

\def\ledd{L_{\rm Edd}}

\def\msun{{\,M_\odot}}

\newcommand\lnet{\Lambda_{\rm net}}
\def\refindent{\par\noindent\hangindent=3pc\hangafter=1 }
\def\aa#1#2#3{\refindent#1, A\&A, #2, #3}
\def\aasup#1#2#3{\refindent#1, A\&AS, #2, #3}

\def\apj#1#2#3{\refindent#1, {\it ApJ}, #2, #3.}
\def\apjlett#1#2#3{\refindent#1, {\it ApJL}, #2, #3.}
\def\apjsup#1#2#3{\refindent#1, ApJS, #2, #3}

\def\mnras#1#2#3{\refindent#1, {\it MNRAS}, #2, #3.}

\def\refbook#1{\refindent#1}

\newcommand\fx{F_{\rm x}}

\newcommand\pg{P_{\rm gas}}

\newcommand\simlt{\lower.5ex\hbox{$\; \buildrel < \over \sim \;$}}
\newcommand\simgt{\lower.5ex\hbox{$\; \buildrel > \over \sim \;$}}

\newcommand\dm{\dot{m}}

\newcommand\ztop{z_{\rm t}}

\newcommand\teff{T_{\rm eff}}
\newcommand\pdim{{\cal P}}
\newcommand\rozanska{R\'o$\dot{\rm z}$a\'nska}
\newcommand\zycki{$\dot{\rm Z}$ycki}

\begin{document}

\title{Thermal Instability and Photoionized X-ray Reflection in Accretion 
Disks}

\author{Sergei Nayakshin\altaffilmark{1}, Demosthenes
Kazanas and Timothy R. Kallman}

\affil{Laboratory for High Energy Astrophysics, NASA/Goddard Space
Flight Center, Code 661, Greenbelt, MD, 20771}


\altaffiltext{1}{National Research Council Associate}


\begin{abstract}
We study the illumination of accretion disks in the vicinity of
compact objects by an overlying X-ray source.  Our approach differs
from previous works of the subject in that we relax the simplifying
assumption of constant gas density used in these studies; instead we
determine the density from hydrostatic balance which is solved simultaneously 
with the ionization balance and the radiative transfer in a
plane-parallel geometry.  We calculate the temperature profile of the
illuminated layer and the reprocessed X-ray spectra for a range of
physical conditions, values of photon index $\Gamma$ for the
illuminating radiation, and the incident and viewing angles.

In accordance with some earlier studies, we find that the
self-consistent density determination makes evident the presence of a
thermal ionization instability well known in the context of quasar
emission line studies. The main effect of this instability is to
prevent the illuminated gas from attaining temperatures at which the
gas is unstable to thermal perturbations.  Thus, in sharp contrast to
the constant density calculations that predict a continuous and rather
smooth variation of the gas temperature in the illuminated material,
we find that the temperature profile consists of several well defined
thermally stable layers. Transitions between these stable layers are
very sharp and can be treated as discontinuities as far as the
reprocessed spectra are concerned. In particular, the uppermost layers
of the X-ray illuminated gas are found to be almost completely ionized
and at the local Compton temperature ($\sim 10^7 - 10^8$ K); at larger
depths, the gas temperature drops abruptly to form a thin layer with
$T\sim 10^6$ K, while at yet larger depths it decreases sharply to the
disk effective temperature.  For a given X-ray spectral index, this
discontinuous temperature structure is governed by just one parameter,
$A$, which characterizes the strength of the gravitational force
relative to the incident X-ray flux.

We find that most of the Fe K$\alpha$ line emission and absorption
edge are produced in the coolest, deepest layers, while the Fe atoms
in the hottest, uppermost layers are generally almost fully ionized,
hence making a negligible contribution to reprocessing features in
$\sim 6.4-10$ keV energy range.  We also find that the Thomson depth
of the top hot layers is pivotal in determining the fraction of the
X-ray flux which penetrates to the deeper cooler layers, thereby
affecting directly the strength of the Fe line, edge and reflection
features.  Due to the interplay of these effects, for $\Gamma \simlt
2$, the equivalent width (EW) of the Fe features decreases
monotonically with the magnitude of the illuminating flux, while the
line centroid energy remains at 6.4 keV. We provide a summary of the
dependence of the reprocessing features in the X-ray reflected spectra
on the gravity parameter $A$, the spectral index $\Gamma$ and other
parameters of the problem.

We emphasis that the results of our self-consistent calculations are
both quantitatively and qualitatively different from those obtained
using the constant density assumption. Therefore, we propose that
future X-ray reflection calculations should always utilize hydrostatic
balance in order to provide a reliable interpretation of X-ray spectra
of AGN and GBHCs.

\end{abstract}


\section{Introduction}

The importance of X-ray reflection off the surface of cold matter in
the spectra of accreting compact objects has been recognized for a
long time.  Basko, Sunyaev \& Titarchuk (1974) discussed the
reprocessing of X-rays from an accreting neutron star incident on the
surface of the adjacent cooler companion, while Guilbert \& Rees
(1988) considered such reprocessing in ``clouds" of cold matter in the
vicinity of AGN. Lightman \& White (1988) studied similar reprocessing
on the surface of a geometrically thin, optically thick accretion
disk, placing the problem within the ``standard'' paradigm for AGN and
Galactic Black Hole Candidates (GBHC) which asserts the presence of
such disks illuminated by an overlying hot corona (e.g., Liang \&
Price 1977; Galeev, Rosner \& Vaiana 1979; Haardt, Maraschi \&
Ghisellini 1994).  The consequences of X-ray reprocessing in this
specific geometry (arrangement) have been explored in detail
theoretically and observationally.

Lightman \& White (1988) and White, Lightman \& Zdziarski (1988)
computed the Green's function for the angle-averaged reflected
spectrum assuming that the illuminated material is cold, non-ionized,
and has a fixed gas density. Soon thereafter, a number of authors
included the effects of ionization on the structure of the iron lines
and the reflected continuum (e.g., George \& Fabian 1991; Turner et
al. 1992). Krolik, Madau \& \zycki (1994) calculated the X-ray
reflection and the iron lines from a putative distant obscuring torus,
while Magdziarz \& Zdziarski (1995) calculated angle-dependent
reflection component off cold matter; Poutanen, Nagendra \& Svensson
(1996) included polarization in their calculations of the reflected
spectra off neutral matter and Blackman (1999) enunciated the
influence of the possible concave geometry of the disk on the iron
line profile.

To date, the most careful (in terms of the ionization physics)
calculations of the X-ray reflection component and the iron lines from
ionized accretion disks in AGN are probably those of Ross and Fabian
(1993), Matt et al. (1993), \zycki et al. (1994), and more recently
for GBHCs of Ross, Fabian \& Brandt (1997) and Ross, Fabian \& Young
(1999).  All these studies made use of a simplifying assumption,
namely that the density in the illuminated gas be constant and equal
to the disk mid-plane value. The justification for this assumption was
the fact that in the simplest version of radiation-dominated
Shakura-Sunyaev disks (Shakura \& Sunyaev 1973, hereafter SS73, see
their \S 2a), thought to be the case in the majority of the observed
sources, the gas density is roughly constant in the vertical
direction.  Note that the accretion disk is radiation-dominated as
long as dimensionless accretion rate, $\dm \equiv \eta \dot{M}
c^2/\ledd \equiv L/\ledd$ ($\ledd$ is the Eddington accretion rate,
$L$ is the disk bolometric luminosity, $\eta=0.06$ is the radiative
efficiency of the standard disk in Newtonian limit), is greater than
about $\sim 5\times 10^{-3}$.

Recently, Ross, Fabian \& Young (1999) suggested that the vertical
density structure may follow a Gaussian law with the vertical
coordinate $z$ ($\rho(z) = \rho(0) \exp[-(z/z_0)^2]$, where $z_0$ is
the scale height, and $\rho(0)$ is the central disk density). They
showed that weakness of the observed Fe features in spectra of some
GBHCs could be interpreted as a result of high degree of ionization of
the exposed gas, as opposed to the cold matter intercepting a smaller
fraction of the emitted X-rays thereby preserving the paradigm of a
cold disk with an overlying hot corona\footnote{Similar suggestions
were made earlier in an unpublished paper by Nayakshin \& Melia
(1997c), and in Nayakshin (1998a,b).}.

However, a self-consistent approach requires that the gas density be
determined from the condition of hydrostatic equilibrium {\em solved
simultaneously} with ionization, energy balance and radiation transfer
equations rather than be assumed. As such, the approximation of the
constant gas density was made by Shakura \& Sunyaev (1973) for disks
heated by viscous dissipation and for large optical depths
only. Therefore, even though the disks we will be considering may be
radiation dominated, the constant density assumption may apply only to
their deep ``cooler" layers, where the X-ray heating is negligible. On
the contrary, in the upper disk layers, this heating exceeds the
viscous heating by orders of magnitude (see \S \ref{sect:xstar}), and
thus the constant density approximation of SS73 cannot be simply
extended to these layers. Further, a self-consistent density
determination is especially important because of the existence of a
thermal instability that allows (and requires under some conditions!)
illuminated adjacent gas regions to have vastly different densities
and temperatures but, by necessity, similar pressure (see Krolik,
McKee \& Tarter 1981 -- hereafter KMT; Raymond 1993; Ko \& Kallman
1994; \rozanska \& Czerny 1996, and \rozanska 1999). For example, when
the gas pressure in the illuminated atmosphere falls below a fraction
of the incident X-ray radiation pressure $\fx/c$ ($\fx$ is the
incident flux), the gas temperature may jump from low values (of the
order $\sim 10^5$ Kelvin for AGN) to the Compton temperature, which is
typically as large as $\sim 10^7 - 10^8$ K. This large jump in
temperature makes it impossible to approximate the gas density by
either a constant or Gaussian profile, as we will see below.

In this paper, we present an X-ray reprocessing calculation that
matches or exceeds the most detailed previous works in terms of the
radiation transfer and ionization balance calculations, but also
relaxes the assumption of the constant gas density. We adopt a
plane-parallel geometry and gravity law appropriate for the standard
geometrically thin SS73 disk, and solve for the gas density via
hydrostatic pressure equilibrium, taking into account the pressure
force due to the X-radiation as well.

The structure of the paper is as follows: In \S \ref{sect:s-curve} we
present general considerations on how the thermal instability may
influence the X-ray reprocessing and review previously known results.
In \S \ref{sect:numerics} we describe the numerical methods we use to
solve the radiation transfer, ionization, energy, and hydrostatic
balance equations. Readers not interested in these details may skip
the latter section and proceed to \S \ref{sect:simple}, where we show
several representative tests and compare our results with those
obtained with the usual constant density
assumption. Radiation-dominated disks are discussed in \S
\ref{sect:rd}. Complications arising due to possible X-ray driven gas
evaporation in the case of accretion disks with magnetic flares are
discussed briefly in \S \ref{sect:evaporation}. We summarize our
results and give our conclusions in \S \ref{sect:summary}. The
Appendix contains notes on the importance of the thermal conduction,
two-phase cloudy structure and the intrinsic disk viscous dissipation
for the given problem.

\section{Thermal Ionization Instability and X-ray Reflection}
\label{sect:s-curve}

In order to simplify our treatment we will assume in what follows that
the geometry of the X-ray illuminated gas is a plane parallel one,
i.e., the vertical $z$-coordinate is the only dimension relevant to
the problem. The surface of the accretion disk is illuminated by
X-rays with a power-law plus exponential roll-over spectrum (the power
law index, $\Gamma= 1.5 - 2.4$, the cutoff energy is fixed at $E_{\rm
cut} = 200$ keV). Note that due to the fact that most of this
radiation is thermalized in the disk and consequently re-radiated, the
ionizing spectrum consists, in the zeroth approximation, of the
incident X-ray spectrum plus the reprocessed black-body radiation with
an equal amount of flux.  The vertical coordinate $z$ is counted from
the disk mid-plane.  The optical depth in the reflecting material is
measured from the top of the illuminated layer, i.e., $\tau(z=\ztop) =
0$, where $\ztop$ is the vertical coordinate of the reflector's top,
to be found in a full self-consistent calculation.

The existence of multiple temperature solutions for an X-ray
illuminated gas at a given pressure was shown by Buff \& McCray (1974)
and then elaborated by KMT. It was noted that some of these solutions
are unstable. This instability conforms to the criterion for thermal
instability discovered by Field (1965). He argued that a physical
system is usually in pressure equilibrium with its surroundings. Thus,
any perturbations of the temperature $T$ and the density $n$ of the
system should occur at a constant pressure.  The system is unstable
when
\begin{equation}
\left({\partial \Lambda_{\rm net}\over \partial T}\right)_{P} < 0,
\label{field}
\end{equation}
where $P$ is the full gas pressure, and the ``cooling function,''
$\lnet$, is the difference between cooling and heating rates per unit
volume (divided by the gas density $n$ squared).

In ionization balance studies, it is convenient to define the
``density ionization parameter'' $\xi$, equal to (e.g., KMT)
\begin{equation}
\xi = {4\pi \fx\over n_H},
\label{xid}
\end{equation}
where $n_H$ is the hydrogen density. For situations in which the
pressure rather than the gas density is of physical importance one
defines the ``pressure ionization parameter'' as
\begin{equation}
\Xi \equiv {\fx\over c\pg}\,,
\label{xip}
\end{equation}
where $\pg$ is the full gas pressure due to neutral atoms, ions and
electrons (we neglect the trapped line radiation pressure in this
paper). KMT showed that the instability criterion (\ref{field}) is
equivalent to
\begin{equation}
\left( {d\Xi\over d T}\right)_{\lnet=0}\, < 0  \; ,
\label{fcond}
\end{equation}
where the derivative is taken with the condition $\lnet = 0$
satisfied, i.e., when the energy and ionization balances are
imposed. In this form, the instability can be readily seen when one
plots temperature $T$ versus $\Xi$, since the unstable parts of the
curve are the ones that have a negative slope.

\begin{figure*}[H]
\centerline{\psfig{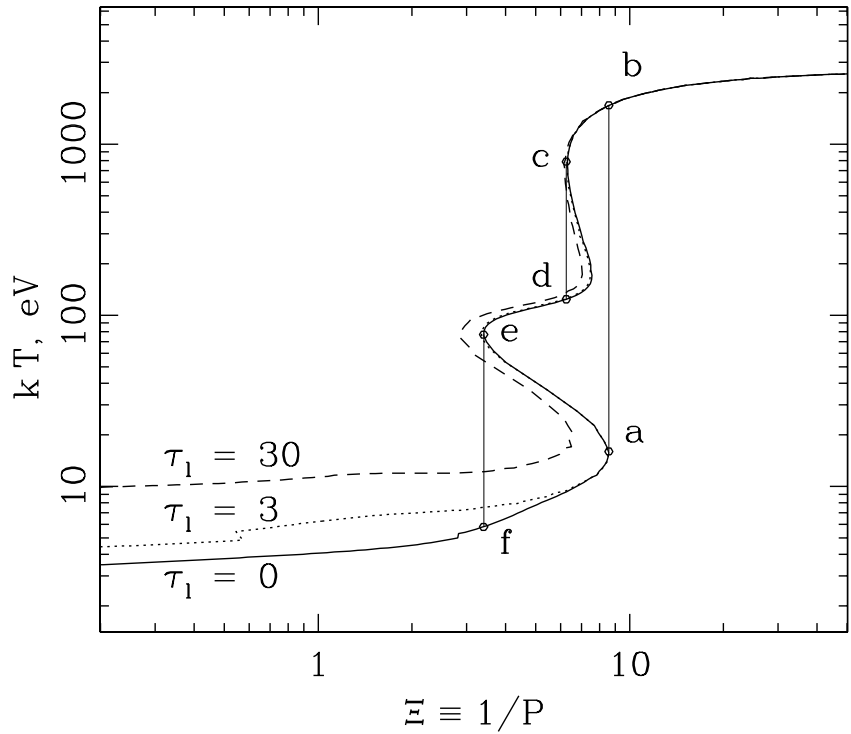}}
\caption{Several illustrative ionization equilibrium curves
(``S-curves'') showing dependence of the gas temperature, $T$, on the
ionization parameter $\Xi$ for three different values of the optical
depth in resonance lines. See text for discussion.}
\label{fig:scurves}
\end{figure*}

In Figure (\ref{fig:scurves}), we show several illustrative ionization
equilibrium curves for the case with the ionizing X-ray flux $\fx =
10^{16}$ erg sec$^{-1}$ cm$^{-2}$, photon spectral index $\Gamma =
1.9$ and an equal amount of the black-body flux with temperature equal
to the effective temperature for the given $\fx$, i.e., $kT= k\teff
\simeq 10$ eV. Since these curves produce a shape somewhat similar to
the letter ``S'', they are commonly referred to as ``S-curves''.  The
solid curve shown in Figure (\ref{fig:scurves}) was computed assuming
that all resonant lines are optically thin, and we will concentrate on
this case for now (significance of the other curves in
Fig. \ref{fig:scurves} will be explained in \S
\ref{sect:multiplicity}).  One can see that equilibria in temperature
range $kT \sim 20 - 90$ eV and $kT \sim 150 - 10^3$ eV are
unstable. It is also worthwhile pointing out that the instability
criterion (eq. [\ref{fcond}]) can be shown to be equivalent to $c_s^2
= d \pg/d\rho < 0$, where $\rho$ is the gas density and $c_s$ is the
sound speed (see Nayakshin 1998c, \S 4.5.) An imaginary value of the
sound speed in the regions with the negative slope of the ionization
balance curve is clearly an ``unusual'' situation, a fact that perhaps
helps to clarify the existence of the thermal instability.

A number of authors have investigated the structure of X-ray
illuminated atmospheres in the case of a star in a binary system
(e.g., Alme \& Wilson 1974; Basko et al. 1974; McCray \& Hatchett
1975; Basko et al. 1977; London, McCray, \& Auer 1981). These authors
concluded that, as one moves from the star's interior to the
atmosphere (that is from $\Xi\ll 1$ to $\Xi=\infty$), the gas
temperature first follows the S-curve until point (a)
[cf. Fig. \ref{fig:scurves}] is reached. For $\Xi > \Xi(a)$, the
photo-ionization heating exceeds the cooling and no equilibrium is
possible until the gas heats up to a fraction of the Compton
temperature, $kT_c\simgt 1$ keV, where even the high-$Z$ elements
become nearly completely ionized. At this state, the gas cooling and
heating is effected predominantly by Compton heating and Compton and
bremsstrahlung cooling (KMT). The transition from point (a) to point
(b) is very sharp\footnote{The line cooling complicates this
situation; we will come back to this question in \S
\ref{sect:multiplicity}}.  In fact, since in the case of a star the
escape energy for a proton is around $E_{es}\sim 100 eV$, and the
Compton temperature $k T_c\sim$~few~keV~$\gg E_{es}$, a strong wind
results. McCray \& Hatchett (1975) treated the temperature
discontinuity as a deflagration wave, in which case the gas pressure
is also discontinuous across the (a-b) boundary; it is the momentum
flux that is continuous across the transition.

We are primarily interested in the inner part of an accretion disk,
where most of the X-rays are produced and where most of the X-ray
reprocessing features presumably arise. Because the gas Compton
temperature, $T_c\simlt 10^8$ K, is smaller than the virial
temperature in the inner disk by several orders of magnitude, no
thermally driven wind will escape from the system, and thus there
should be a static configuration for the X-ray heated layer in this
region (see Begelman, McKee \& Shields 1983). Thus, the problem of
finding the gas density is reduced to finding its hydrostatic equilibrium 
configuration. 

Raymond (1993) and Ko \& Kallman (1994) have calculated the structure
of the accretion disk atmospheres around LMXBs at large radii with
special attention to the unknown origin of the UV and soft X-ray
emission lines. The illuminating spectrum was assumed to be thermal
bremsstrahlung with $T = 10^8$ K. The work of these authors was
pioneering in that they calculated the density of the illuminated gas
using hydrostatic balance, thus taking the thermal instability into
account consistently.  \rozanska \& Czerny (1996), \rozanska (1999)
considered the structure of X-ray illuminated disk atmospheres in AGN,
including the effects of thermal conduction. The radiation transfer
methods used by these last authors did not allow them to study the
effects of spectral reprocessing in detail.

Our paper is a natural extension of the work of these authors.  Since
we study accretion disks in Seyfert 1 Galaxies and GBHCs rather than
in LMXBs as did Raymond (1993) and Ko \& Kallman (1994), we use the
ionizing spectrum typical for Seyfert 1 Galaxies and GBHCs and deal
with the innermost part of the accretion disk. Compared with work of
\rozanska \& Czerny (1996) and \rozanska (1999), we solve the radiation
transfer ``exactly'' by using the exact frequency dependent cross
sections and the variable Eddington factor method to take into account
anisotropy of the radiation field.

\section{The Numerical Approach}\label{sect:numerics}

\subsection{Radiation Transfer}
\label{sect:rad_transfer}

Here we describe the part of our code that performs the radiation
transfer. The incident X-rays come with a given specific intensity
$I_x(E, \mu)$, where $E$ is the photon energy and $\mu$ is the cosine
of the angle that the given direction makes with the normal to the
surface of the disk (note that $I_x(E, \mu)$ is the intensity already
integrated over the azimuthal angle, i.e., it is $2\pi$ times the
usual radiation intensity). The X-ray intensity is normalized in such
a way as to give the total X-ray flux incident on the reflector, $\fx$
(erg cm$^{-2}$ sec$^{-1}$), integrated over all photon energies and
all incident angles:
\begin{equation}
F_x = \int_{-1}^0 |\mu| \, d\mu\, \int_0^{\infty} dE\, I_x(E, \mu)\; .
\label{fx}
\end{equation}
We limit ourselves to treating only $\tau_{\rm max} = 4$ upper Thomson
depths of the illuminated layers, since it is both sufficient and
practical (see \S \ref{sect:xstar}). We break the reflecting layer
into $N_z$ (typically a few hundred) individual bins. The binning is
the finest around the discontinuity in the gas temperature and
density.

The ultimate goal of the radiation transfer is to find the photon
intensity, $I(E, \mu, \tau_t)$ as a function of photon energy $E$,
angle (or $\mu$) and position (equivalently $\tau_t$) everywhere in
the slab. Luckily, the ionization balance part of the calculations
only depends on the intensity integrated over all angles, $J(E,
\tau_t)$:
\begin{equation}
J(E, \tau_t) = \int_{-1}^{+1} \, d\mu I(E, \mu, \tau_t)\;,
\label{jnu}
\end{equation}
which then allows us to use the variable Eddington factors method (see
Mihalas 1978, \S 6.3). The variable Eddington factors $g(E, \tau_t)$
are defined as the ratio of the second to the zeroth moment of the
photon intensity: $g(E, \tau_t) \equiv K(E, \tau_t)/J(E, \tau_t)$,
where $K(E, \tau_t)$ is given by an equation analogous to Equation
(\ref{jnu}), but with the additional factor of $\mu^2$ in the
integral. Mihalas (1978) shows that the transfer equation then reduces
to
\begin{equation}
{\partial^2\over\partial \tau^2}\,\left[ g(E, \tau_t)\, J(E,
\tau_t)\right] = J - S \;,
\label{eddap}
\end{equation}
where $S(E, \tau_t)$ is the source function, defined as
\begin{equation}
S(E, \tau_t) \equiv {\alpha_s\over \alpha_s + \alpha_a}
\; C J(E, \tau_t) \; + \; {1\over \alpha_s + \alpha_a}
\; j(E, \tau_t)\;,
\label{sf}
\end{equation}
where $\alpha_s$ and $\alpha_a$ are the scattering and absorption
coefficients, respectively: $\alpha_s = n_e \sigma_{\rm kn}(E)$, where
$n_e = 1.2 n_H$ is the electron density, $\sigma_{\rm kn}(E)$ is the
Klein-Nishina cross section; $\alpha_a = n_H \sigma_a(E)$, where $n_H$
is the hydrogen density and $\sigma_a(E)$ is the continuum absorption
cross section calculated by XSTAR (Kallman \& McCray 1982; Kallman \&
Krolik 1986).  Further, $d\tau$ in Equation (\ref{eddap}) is a
differential of the total optical depth, so that it is a function of
both photon energy and the location in the slab ($\tau_t$) and is
given by $d\tau = (\alpha_s + \alpha_a) d z$ where $d z$ is the extent
of the given bin in the $z$-direction.  Finally, $j(E, \tau_t)$ is the
local gas emissivity integrated over all angles, and $C$ is the
Compton scattering operator, discussed below.

The Compton scattering is treated similarly to the Ross \& Fabian
(1993) treatment of Compton down-scattering of incident X-rays: the
scattered photons are assumed to be distributed according to a
Gaussian profile, $P(E_0\rightarrow E)$ (normalized to unity),
centered at
\begin{equation}
E_c = E_0 \left (1 + 4 \theta - \varepsilon_0\right )\;,
\label{ec}
\end{equation}
where $\theta\equiv kT/m_e c^2$ is the dimensionless electron
temperature, $E_0$ is the initial photon energy and
$\varepsilon_0\equiv E_0/m_ec^2$. The energy dispersion of the
Gaussian is
\begin{equation}
\sigma = \varepsilon_0 \left( 2\theta + 
(2/5)\, \varepsilon_0^2\right)^{1/2}
\label{ed}
\end{equation}
As shown by Ross \& Fabian (1993), and as we found in the course of
our numerical experimentation, this treatment describes adequately the
down-scattering of photons of energy less than $\sim 200$ keV.

Ross \& Fabian (1993) used a modified Kompaneets equation to describe
Compton scattering of the low energy photons. We will see in \S
\ref{sect:hydro_balance} that an accurate numerical flux conservation
throughout the reflecting layer is a necessary condition for solving
the pressure balance satisfactorily. For best numerical flux
conservation, we found that it is preferable to treat all the photons
in the same way, avoiding division of the photon intensity on the
incident and the `diffuse' components. Therefore, we use the same
approach (i.e., the Gaussian emission profile) for Compton scattering
of photons of all energies. This is permissible, since for most of our
applications below, the average photon with $E\ll m_e c^2$ gains or
loses little energy ($|\Delta E|/E\sim 4\tau_t^2\theta\ll 1$) before
it escapes from the layer, and thus Compton scattering is essentially
monochromatic. We conducted several tests of Comptonization of
different photon spectra by slabs of different temperatures (in the
range $T= 10^5$ -- $10^8$ K, the typical values for our problem) via
both techniques (i.e., the Gaussian emission profile and the
Kompaneets equation) and found no noticeable differences in the
results for the slabs of Thomson depth $\tau_t\sim$ few (but see the
end of \S \ref{sect:summary}). Accordingly, the result of the Compton
scattering operator acting on $J(E, \tau_t)$ in our approach is
\begin{equation}
CJ(E, \tau_t) \equiv - \, J(E, \tau_t) + J'(E, \tau_t) \;,
\label{cop}
\end{equation}
where $J'(E, \tau_t)$ is the intensity of once scattered photons found
by convolving the Gaussian re-distribution profile $P(E'\rightarrow
E)$ with the
unscattered intensity $J(E, \tau_t)$:
\begin{equation}
J'(E, \tau_t) \equiv \, {1\over \sigma_{\rm kn}(E)} \, \int dE' J(E',
\tau_t) P(E'\rightarrow E) \sigma_{\rm kn}(E')\;,
\label{jprime}
\end{equation}

The radiation transfer Equation (\ref{eddap}) is supplemented by
boundary conditions. On the top of the illuminated layer (i.e., for
$\tau_t = 0$), we require that the down propagating flux is equal to
the incident X-ray flux:
\begin{equation}
h(E, 0) J(E, 0) - {\partial\over \partial \tau}\left[g(E,0)
J(E,0)\right] = F_x(E)\;,
\label{bct}
\end{equation}
where $h(E, 0)$ is the ratio of the first to the zeroth moment of the
intensity at $\tau_t = 0$ (see Mihalas 1978, \S 6.3, especially text
after Equation (6-42)).

We assume that all of the incident X-ray flux is reprocessed into the
thermal disk flux deep inside the disk (e.g., Sincell \& Krolik 1997),
and that there is also an intrinsic disk flux $F_d$ whose magnitude is
simply given by the formulae of SS73. We found that
the most physical and numerically accurate boundary condition is a
mirror boundary condition for $J(E, \tau_{\rm max})$, similar to 
that used by Alme \& Wilson (1974), with the appropriate correction 
due to the additional disk flux $F_d$:
\begin{equation}
{\partial\over \partial \tau}\left[g(E,\tau_{\rm max}) J(E,\tau_{\rm
max})\right] = F_d(E)\;.
\label{bct1}
\end{equation}
If $F_d(E)=0$, then equation \ref{bct1} leads to the exact mirror
boundary condition, i.e., every photon is reflected without any change
in its energy. This is somewhat unphysical because in reality, due to
photo-absorption, X-ray albedo is less than unity, but since only a
tiny fraction of X-rays will reach the bottom of the reprocessing
layer, we are committing a small error.

The variable Eddington factors are obtained in the usual way (Mihalas
1978, \S 6.3). One defines the function $u(E, \mu, \tau_t)$ as $u(E,
\mu, \tau_t)\equiv [I(E, \mu, \tau_t) + I(E, -\mu, \tau_t)]/2$ for
$\mu > 0$. The angle dependent equation of the radiation transfer
can then be written as 
\begin{equation}
\mu^2\,{\partial^2 \over\partial \tau^2}u(E, \mu, \tau_t)\, = J - S \;,
\label{rtex}
\end{equation}
and it can be solved {\em exactly} once $J(E, \tau_t)$ and the source
function, $S(E, \tau_t)$, are known. The solution of this equation
allows us to refine our initial estimate for the variable Eddington
factors $g(E, \tau_t)$ and the flux factors $h(E, \tau_t)$. These new
functions are then employed in Equation (\ref{eddap}) to obtain the
next iteration of $J(E, \tau_t)$. We repeat this cycle without calling
XSTAR (i.e., with fixed opacities and emissivities obtained from the
last XSTAR call) typically 30 times such that the Eddington factors
converge to a good degree.

\subsection{Photoionization Calculations}
\label{sect:xstar}

Our treatment of the thermal and ionization balance is closest to the
work of \zycki et al. (1994). As mentioned earlier, we use the
photoionization code XSTAR (Kallman \& McCray 1982; Kallman \& Krolik
1986). We start from the first zone situated on the top of the
reflecting gas and work our way to the bottom of the layer. We provide
XSTAR with the ionizing intensity, $J(E, \tau_t)$, for each zone
computed in the previous iterations or given by the initial value for
the first iteration. The resonance line optical depth for photons
emitted in the first layer is assumed to be zero for each line
included in XSTAR. The latter then provides us with line opacities for
this zone, which we use to compute the line optical depths for the
line photons emitted in the next zone. We repeat this process for
every spatial bin; therefore, the optical depth of a particular
resonant line is a sum of the appropriate line depths for all the
previous zones above the given one.

The treatment of the coldest part of the reflecting layer deserves
special attention. In terms of the number of the included elements,
ionization states and excited states XSTAR is one of the most
extensive codes existing at the present time. However, the code does
not include stimulated emission or de-excitation processes of certain
excited states, so it may incorrectly estimate the cooling rate under
some conditions. As in Raymond (1993) and \zycki et al. (1994), we set
the gas temperature to the effective one ($T_{\rm eff}$), if the value
computed by XSTAR is lower than $T_{\rm eff}$.  Clearly, this
approximation is not very accurate for photon energy $\sim$ few
$\times kT_{\rm eff}$, and may not give correct opacities and
emissivities in that energy range.  The focus of this paper, however,
is on the X-ray reflection feature and the iron line, i.e. the photon
energy range $ E \simgt 1$ keV, so that the approximation described
above should suffice.

Further, in most previous studies, one usually sets a constant
black-body or modified black-body continuum to propagate in the upward
direction (e.g., Ross \& Fabian 1993; Ko \& Kallman 1994; \zycki et
al. 1994). Since we intend to include the radiation pressure in the
hydrostatic balance equation, this approach is not suitable for us. We
need to propagate the quasi-thermal component through the layer just
as we do the X-rays.  The difficulty with this is that we previously
introduced the minimum temperature $T=T_{\rm eff}$, which forbids
XSTAR to solve the energy balance equation if it leads to $T < T_{\rm
eff}$. Therefore, the opacities and emissivities obtained from XSTAR
in this way will not match each other (i.e., total energy absorbed
$\neq$ total energy emitted). To cope with this problem, we redefine
the emissivities obtained from XSTAR in this regime in the following
manner: when $T=T_{\rm eff}$, we count the line emission only if the
line energy $E > 100$ eV, otherwise we set it to zero (because we are
mostly interested in the X-ray emission). The continuum emission is
assumed to be given by $j(E, \tau_t) = \alpha_a B(E, T_{\rm eff})$,
where $B(E, T_{\rm eff})$ is the Planck function at $T=T_{\rm
eff}$. This choice of $j(E, \tau_t)$ is appropriate since it is
exactly the Kirchoff-Planck law and is often adopted as an expression
for the rate of thermal emission. The quasi-thermal emission spectrum
$j(E, \tau_t)$ is renormalized in such a way as to satisfy the energy
equilibrium equation.

Finally, one practically important question is how deep in the
illuminating layer should we carry our calculations (i.e., how large
should $\tau_{\rm max}$ be?). In terms of the energy balance for the
illuminated layer, it is sensible to extend the computational domain
down to the Thomson ``viscous heating depth'', $\tau_{vh}$, such that
for $\tau_t > \tau_{vh}$ the intrinsic disk viscous heating wins over
the X-ray heating. To estimate value of $\tau_{vh}$, we proceed in the
following manner. The intrinsic disk heating per hydrogen atom, in the
framework of SS73 model, is $Q^+_d = F_d/(n_H(0) H)$, where $n_H(0)$
is the mid-plane hydrogen density.  Further, we follow Sincell \&
Krolik (1997), who derived a useful approximate expression for the
X-ray heating rate (see their equation 32).  Re-writing it in terms of
Thomson depth in the limit $\tau_t \gg 1$,
\begin{equation}
Q^+_x\sim 0.33 \, \sigma_t \fx\, \left[{ \exp(-\tau_t/22)\over
\tau_t}\right]\;.
\label{qx}
\end{equation}
Hence, the ratio of the X-ray heating rate to viscous dissipation
is
\begin{equation}
{Q^+_x\over Q^+_d} \sim 0.33\, {\fx\over F_d}\,
{\tau_d\over \tau_t} \; \exp(-\tau_t/22) \, ,
\label{qratio}
\end{equation}
where $\tau_d$ is the total Thomson depth of the accretion
disk. Clearly, the X-ray heating exceeds the viscous heating by orders
of magnitude for $\tau_t \simlt 20$, unless $\fx\ll F_d$, a situation
of no particular interest to us. Accordingly, we neglect the viscous
heating in the illuminated layer (although see the Appendix).

Further, it is computationally impractical to extend computational
domain to $\tau_t \sim 20$, since the computing resources are limited.
Fortunately, very few X-rays penetrate to Thomson depth greater than
$\tau_t\simgt$ few. Therefore, we choose to treat with the full
formalism developed in this paper only the upper 4 Thomson depths. In
addition, in order to reproduce the thermal component in the reflected
spectra, one ought to have some material at that temperature (since
our mirror boundary condition does not by itself create thermal flux
in the limit $\fx\gg F_d$). In the realistic situation, this material
lies below $\tau_{vh}\simgt 20$, according to the foregoing
discussion.  Since we limit ourselves to $\tau_{\rm max}=4$, we
instead require that the gas temperature always be the effective
temperature for $3< \tau_t \leq 4$. Via numerical experimentation, we
found that the Thomson depth of $\tau_t= 3$ is indeed adequate to
compute the reprocessing features, and that the region $3< \tau_t \leq
4$ is sufficiently optically thick to produce the thermalized part of
the reflected spectrum even if $F_d=0$. These restriction on $\tau_t$
does not affect our results at all. As we found (see fig.
\ref{fig:temper1}) our highest ionization case becomes ``cold" below
$\tau_t \simeq 2$. As apparent in fig. (\ref{fig:spectra1}) in this
specific case the associated reprocessing features are extremely weak.
Therefore, in a situation in which the ionized layer extends to
$\tau_t > 3$ such features would be completely unobservable.

\subsection{Hydrostatic balance}
\label{sect:hydro_balance}

Currently, one of the most widely considered models of the X-ray
emission from accretion disks is that where X-rays come from localized
active regions, thought to be magnetic flares (e.g., Galeev, Rosner \&
Vaiana 1979; Haardt, Maraschi \& Ghisellini 1994; Svensson 1996;
Nayakshin 1998a,b,c). In the framework of that model, the X-ray flux
from the active region is very much larger than the radiation flux
generated within the disk. Nayakshin \& Melia (1997a), Nayakshin
(1998b) and Nayakshin \& Dove (1999) found that the radiation pressure
from the X-rays of magnetic flares may be larger than the unperturbed
gas pressure of the accretion disk atmosphere.  Thus, we ought to
incorporate the effects of the incident radiation pressure in our
pressure balance calculations for generality.

The radiation force per hydrogen atom inside the reflecting layer,
${\cal F}$, is equal to
\begin{eqnarray}
{\cal F}(\tau_t) = n_H \int_{-1}^{+1} \mu \, d\mu \, \int_0^{\infty}
d(E/c)\, I(\mu, E, \tau_t) \sigma(E, \tau_t) \nonumber \\ \;= (F_x/c)
n_H \Delta \sigma(\tau_t) \;,
\label{frad}
\end{eqnarray}
where we introduced $\Delta \sigma(\tau_t)\equiv F_x^{-1}\int \mu d\mu
\int d E I \sigma$, which has dimensions of a cross section ($\sigma$
here is the total cross section per hydrogen atom).  Note that this
quantity can be positive as well as negative; a positive $\Delta
\sigma(\tau_t)$ corresponds to radiation force pointing upward, which
would happen when the cross section for the interaction of matter with
the reprocessed emission is larger than that with the X-rays.  The
equation for the pressure balance now reads
\begin{equation}
{\partial \pg\over\partial z} = - {G M\rho\over R^2}\, {z\over R} \, +
\, {F_x \Delta\sigma(\tau_t) n_H(z)\over c} + {F_d\sigma_T n_H\over
c}\; ,
\label{heq1}
\end{equation}
It is convenient to re-write this equation in terms of dimensionless
quantities that will figure prominently in the ionization and
radiation transfer calculations. These are: the gas pressure
normalized by the incident X-ray radiation pressure, ${\cal P}\equiv
c\pg/F_x$; the differential Thomson depth, $d\tau_t\equiv -n_e(z)
\sigma_t dz$; and the compactness parameter $l_x$ defined as
\begin{equation}
l_x\equiv {\sigma_t F_x H\over m_e c^3}\;.
\label{lx}
\end{equation}
We will further introduce a dimensionless ``gravity parameter'', $A$,
for brevity of notations:
\begin{equation}
A \equiv {R_g \mu_m\over 2 R m_e} \left({H\over R}\right)^2
l_x^{-1}\;,
\label{adef}
\end{equation}
where $\mu_m$ is defined by $\mu_m = \rho/n_H$ and $H$ is the scale 
height of the SS73 accretion disk. The parameter $A$
characterizes the strength of the gravity term relative to that of the
X-radiation pressure in Equation (\ref{heq1}). The hydrostatic balance
equation is given by
\begin{equation}
{\partial {\cal P}\over\partial \tau_H} = \,\left[A\,{z\over H} \, -
\, {\Delta\sigma \over \sigma_t} - {F_d\over \fx}\right]\;,
\label{heq2}
\end{equation}
where $d\tau_H\equiv - n_H \sigma_T dz = (n_H/n_e) d\tau_t$.

\subsection{Pressure boundary conditions}\label{sect:pbc}

The solution of the hydrostatic balance equation (\ref{heq2}) is
subject to boundary conditions. The boundary condition on the top of
the illuminated layer is $\pg(z=\ztop) = 0$, obviously (although note
that if a hot corona is placed on the top of the illuminated layer,
then $\pg(z=\ztop) \neq 0$ -- see \rozanska et al. 1999). Practically,
we set $\pg(\ztop)$ to a small fraction of the incident X-radiation
ram pressure, i.e., ${\cal P} = 10^{-2} - 10^{-3}$. The boundary
condition on the bottom of the illuminated layer requires that the gas
pressure be continuous across the interface between the illuminated
layer and the accretion disk.  Once the pressure profile in the disk
is known, it is a relatively simple matter to iterate on the {\it a
priori} unknown location of the top of the illuminated layer $\ztop$
in order to adjust the gas pressure on the bottom to match the disk
pressure. However, one inherent problem here is that the vertical
structure of the accretion disks is not necessarily well known (see,
e.g., De Kool \& Wickramasinghe 1999). For the gas-dominated accretion
disks, we will simply assume that the density profile follows
exponential law\footnote{Of course, this density profile is applicable
only deep within the disk, where the X-ray heating is negligible,
i.e., below $\tau_t = 4$ by our choice.}:
\begin{equation}
\rho(z) = \rho(0) \exp[- (z/z_d)^2]\;,
\label{rhogd}
\end{equation}
where $z_d$ is the ``average'' scale height, which we approximate as
$z_d^2 = (2 R^3 k_B/GM\mu_m)\, [T_d \teff]^{1/2}$, where $k_B$ is the
Boltzmann constant and $T_d$ is the disk mid-plane temperature. In
principle, one could resolve the vertical structure of the disk to a
``better'' precision by integrating the hydrostatic balance equation
from the disk mid-plane up. However, the distribution of the disk
viscous dissipation with height is not known (SS73 simply assumed it
is proportional to the gas density), precluding the accurate
determination of the temperature profile $T(z)$ of the disk.
Fortunately, as we will see later, this does not bring any serious
complications except for a slight uncertainty in the exact value of
$z_b$.

We are less lucky in the radiation-dominated (RD) case, since there
the boundary conditions are of a greater importance. To see this,
consider the case in which the illuminated layer is completely
ionized, so that the only source of opacity is the Thomson one. In
this case, $\Delta \sigma =0$, and the second term on the right in
equations (\ref{heq1}) and (\ref{heq2}) is zero. Further, note that
without the second term, equation (\ref{heq1}) is exactly the equation
for hydrostatic balance inside the disk. In the RD case, the gas
pressure is negligible everywhere except for very near the edge of the
disk. In other words, inside the disk, the gravity term is canceled by
the intrinsic radiation pressure term (the last term in eq. \ref{heq1}
on the right). As we found, the illuminated layer does not extend
very high in the RD case, e.g., $(z_t-z_b)/H \ll 1$, so that the first
term on the right in eq. (\ref{heq1}) is again almost completely
canceled by the last term. In other words, the net force is the
difference of two very large terms, and hence it is rather sensitive
to the location of the illuminated layer lower boundary $z_b$.

Another serious problem is that the gas density profile in an RD disk
is uncertain to a much greater degree than in the case of a
gas-dominated (GD) disk. As shown in SS73, if one completely neglects
the gas pressure, then the density is constant with height all the way
to $z=H$ in RD disks.  In order to significantly improve the SS73
approximation to the vertical disk structure, one needs to solve the
energy balance equation for the accretion disk {\em simultaneously}
with the pressure balance. This is a more difficult problem (and more
model dependent) than the one we are attempting to solve in this paper
(e.g., De Kool \& Wickramasinghe 1999), and for this reason it is
clearly outside of the scope of our present work.
Therefore, at present, we  make the {\em total}
pressure continuous across the lower boundary and use the
approximation of SS73 for the vertical pressure structure. Since we
require the gas temperature to be equal to the disk effective
temperature and since the radiation field incident on the lower
boundary from the illuminated layer is quite close to the Planckian
function, the radiation pressure is actually continuous through the
boundary by our choice (see discussion below eqs. \ref{bct} and
\ref{bct1}). Thus, if we match the total pressure on both sides of the
boundary to within a very small difference, the gas pressure and
density will also be continuous.

\subsection{The Complete Iteration Procedure}\label{sect:itter}

In order to achieve  static solutions for all the quantities in all
vertical zones, we make the following steps:

1) First, we guess the location of the upper boundary of the
illuminated gas $\ztop$, and we also assume an initial intensity $J(E,
z)$ in all the spatial bins, usually simply equal to the sum of the
incident X-ray intensity and the black body spectrum, and use it as an
input to XSTAR in order to obtain the opacities.  The initial value of
the Eddington variable factors is assumed to be the isotropic value,
i.e., $1/\sqrt{3}$. Then, starting from the top of the illuminated
layer, we solve for the gas pressure in the next zone using
eq. (\ref{heq2}). Given that value of $\pg$, we call XSTAR with the
initial illuminating spectrum. XSTAR provides on output the gas
temperature, density, opacities, line and continuum emissivities for
this zone. We repeat this step for the next zone (i.e., again solve
for its pressure and then call XSTAR), until we reach the bottom of
the transition layer.

2) Having found the gas opacities and emissivities, we proceed to
compute the radiation transfer as described in \S
\ref{sect:rad_transfer}. At this step, we fix the opacities and
emissivities at the value given by the previous iteration of XSTAR,
and make 20-25 iterations of the radiation transfer calculations (made
by an implicit differencing scheme) until the reflected spectrum
converges. The variable Eddington factors are also continuously
updated during this step.

3) At the same time, we check the consistency of the lower pressure
boundary condition. In particular, we compare the pressure at the base
of the illuminated layer (i.e., at $z=z_b$) with the SS73 pressure at the
same height. If the latter is larger than the former, we increase the
value of $\ztop$ from the previous iteration; if not, we decrease it.

4) Now that we have the refined value of $\ztop$, and the new ionizing
spectrum in each bin, we repeat step (1) using the new values for
$\ztop$, the ionizing spectrum and the variable Eddington factors.

We continue to repeat these steps until a convergence is reached. We
found that the best indicator of convergence is the matching of the
boundary pressure at the base of the layer, $P_b$ to the SS73 pressure
at that height $P_{ss}$. We typically require that $|P_b - P_{ss}|
\simlt 0.01 - 0.001$. We also check that the temperature structure of
the illuminated layer stops evolving to the same level of
precision. Typically, we need to repeat the whole cycle 20 or more
times, which translates into $\sim$ a day of CPU of a 500 MHz PC for
about 250 spatial zones and 400 photon energy points.

\subsection{On the uniqueness of a hydrostatic balance solution}
\label{sect:multiplicity}

Figure (\ref{fig:scurves}) shows that due to the thermal instability,
there are two, and sometimes even three stable solutions of the
thermal and ionization balance equations for a given gas pressure and
local ionizing spectrum when the gas pressure is in the range $\pdim
\simeq 0.1 - 0.5$ (equivalently, $\Xi\simeq 2-10$). Thus, one {\em has to
choose} a solution out of the several possible based on additional
considerations. 

\begin{figure*}[t]
\centerline{\psfig{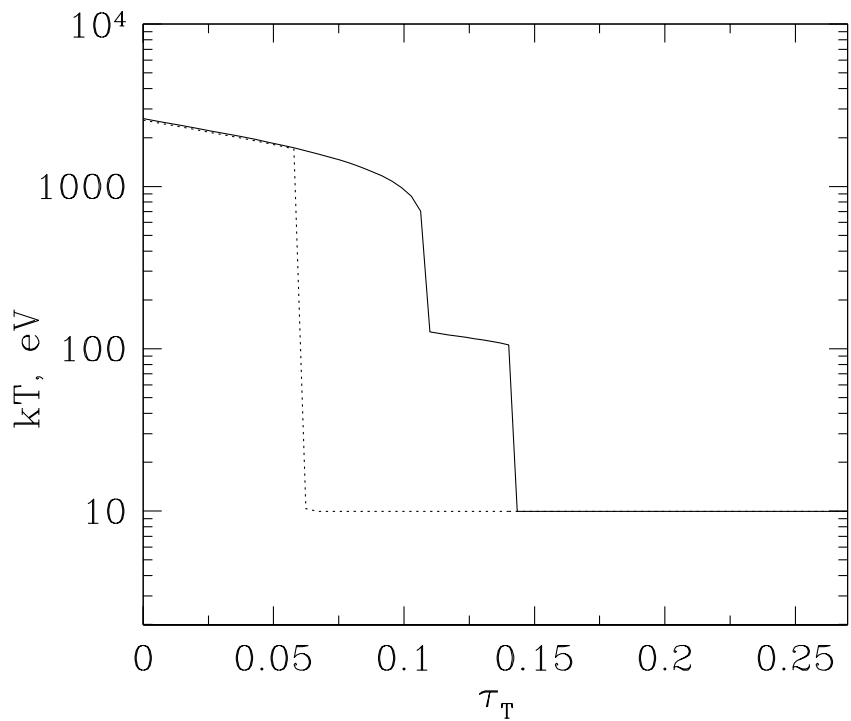}}
\caption{Temperature of the illuminated layer as a function of the
Thomson depth when all the lines are assumed to be optically thin.
The solid curve is calculated with a high initial temperature guess
($k T_{in} = 3$ keV), whereas the dotted curve shows results of a
simulation with a low initial temperature guess ($k T_{in} = 10$
eV). Note that all other parameters in these two cases are equal, so
that the different outcomes illustrate the uncertainty that exists due
to multiple solutions allowed in the thermally unstable region.}
\label{fig:thin}
\end{figure*}

The approach that would leave no ambiguities would be to supplement
our ionization and radiation transfer methods with dynamical equations
for the gas motions in the illuminated layer (e.g., Alme \& Wilson
1974). Such an approach would delineate the different solutions
automatically, since then one would follow the evolution of the gas
density explicitly. For a given density (unlike given pressure), there
is always a thermally stable, unique solution of the ionization and
thermal balance equations (this can be seen from an always positive
slope of the $T$ versus $\xi$ curve; see, e.g., Nayakshin 1998c,
Fig. 5.2). In addition, inclusion of thermal conduction (e.g.,
\rozanska 1999) can also forbid some of the otherwise stable solutions
(see below).

Unfortunately, the numerical complexity of such an approach is 
prohibitive.  The problem with adding the heat conduction to our code
is that in our iterative process, we move from a known temperature in
zone $j$ to solve for temperature in zone $j+1$, whereas thermal
conduction requires knowledge of temperature in zone $j+2$ at this
step. Writing a time-dependent code that would self-consistently treat
the line cooling is also challenging and beyond the scope of the present
work.

Given this, we should be able to find a way to single out a solution
with some desirable property (e.g., within a given temperature range)
out of several possible solutions via our numerical methods.  We
discovered that the easiest way to accomplish this is to give XSTAR an
initial temperature guess, $T_{in}$ that is somewhat close to the
temperature of the solution that we want to pick. This is due to the
fact that XSTAR defines the temperature range where it will seek a
solution in some interval around the initial temperature $T_{in}$. If
there indeed exists a solution that is close to $T_{in}$, then XSTAR
converges to that solution; if that solution does not exist, then
XSTAR expands the interval until it finds a solution.

To avoid complications due to line cooling, we first demonstrate this 
method of separating the solutions with the example of
a calculation in which all the resonant lines were assumed to be
optically thin, i.e., $\tau_l\equiv 0$ for all lines in all spatial
zones. Figure (\ref{fig:thin}) shows the gas temperature as a function
of the Thomson optical depth for two tests. In both tests, the X-ray
flux with $\fx = 10^{16}$ erg s$^{-1}$ cm$^{-2}$ is incident in the direction
normal to the reflecting layer.  Other parameters have the following
values: $\Gamma = 1.9$, $A= 1.2$, and $l_x= 0.01$ (given that, one can
compute the disk vertical height scale $H$; for simplicity, we assume
that the bottom of the reflecting layer is located at $z_b = H$ for
the tests in this section; this does not affect any of our conclusions
here). The solid curve was computed for the case when an initial
temperature guess for each zone is equal to $kT = 3$ keV. This method
ensures that whenever there are multiple solutions, the hottest one is
chosen. In terms of the optically thin S-curve shown in Figure
(\ref{fig:scurves}), this simulation yields the transitions from the
hot Compton-heated branch to the medium stable branch via line (c-d),
and then via line (e-f).  The dotted curve shows results of a test
when an initial gas temperature is close to $\teff$. As we found, this
method always leads to a solution with the smallest possible
temperature, which corresponds to the discontinuity occurring via line
(a-b) in Figure (\ref{fig:scurves}).

\begin{figure*}[t]
\centerline{\psfig{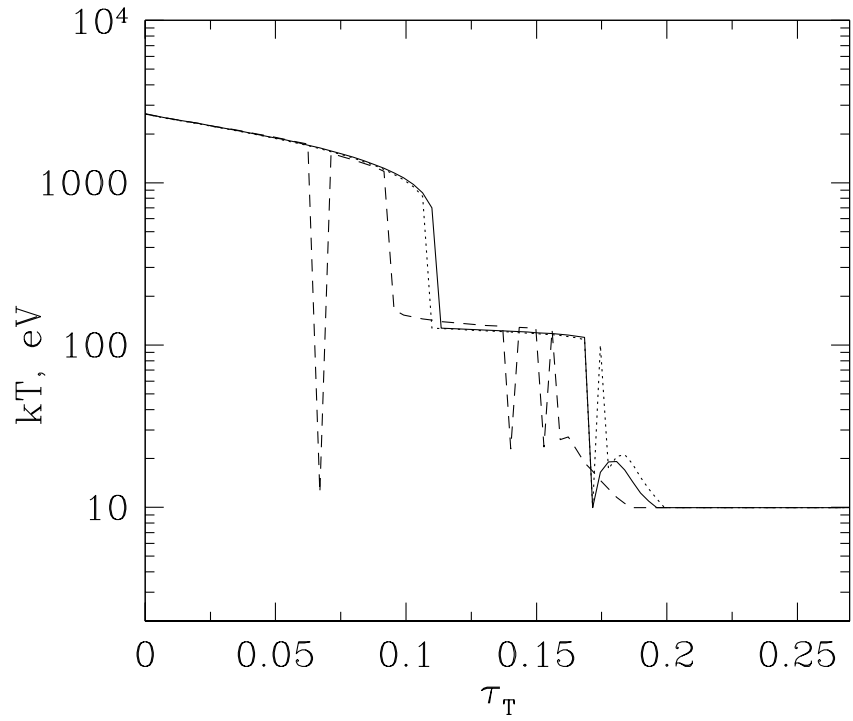}}
\caption{Temperature of the illuminated layer as a function of the
Thomson depth when the line depths are calculated self-consistently.
The dotted curve is for solution found by providing XSTAR with a high
initial temperature guess for each zone ($k T_{in} = 3$ keV); the
dashed curve is for the simulation with a low initial temperature
guess ($k T_{in} = 10 eV$), and the solid curve is computed with
$T_{in} = 1.5\times T_{\rm prev}$, where $T_{\rm prev}$ is the
temperature of the previous zone.}
\label{fig:thick}
\end{figure*}

In Figure (\ref{fig:thick}) we show three tests for the same parameter
values, but with the resonance line optical depths self-consistently
calculated as described in \S \ref{sect:xstar}.  The long-dashed curve
shows the results of a calculation with the initial temperature for
each bin $T\simeq \teff$.  As with the dotted curve of
Fig. (\ref{fig:thin}), the cold solution is recovered rather quickly
at $\tau_t\simeq 0.07$, as soon as it becomes available. Contrary to
the optically thin case shown in Fig. (\ref{fig:thin}), however, the
next zone jumps back to the hot solution. The reason for this is line
cooling.  For a given bin, the line optical depths are calculated by
summing over the line depths of all the previous bins. Since the
material is highly ionized for all zones preceding the one with
$T\simeq \teff$, there is little line optical depth for this zone, and
thus we can use the S-curve generated for optically thin lines (solid
curve in Fig. \ref{fig:scurves}). However, for certain lines the
optical depth of the first cold zone turns out to be significant, and
therefore cooling for all the material with $\tau_t \gtrsim 0.07$
cannot be assumed to be (line-) optically thin anymore.  We then need
to refer to the other two curves in Fig. (\ref{fig:scurves}) that were
calculated for a non-zero line optical depth $\tau_l$ (where it was
assumed for simplicity that all the lines have a given value of
$\tau_l$). In particular, the longed-dashed curve in Figure
(\ref{fig:scurves}), shows that there is no cold solution with
$\Xi\simeq \Xi_b$, where $\Xi_b$ is the pressure ionization parameter
at point (b).  The only acceptable solution is the hot solution, which
then explains why the temperature in the next zone jumps back to the
Compton equilibrium. Note that a finer (than the one we used) zoning
around this point will not remove this behavior, since this behavior
is physical rather than numerical. The two other dips in temperature
around $\tau_T\simeq 0.135$ and $\tau_T\simeq 0.155$ in the dashed
curve in Fig. (\ref{fig:thick}) are caused by similar effects due to
different lines.

Since the sharp dips in the temperature will cause a strong conductive
heating, the cold temperature guess method seems to be inappropriate.
Further, as discussed by London et al. (1981), the effects of the line
cooling may lead to a non-steady behavior in the illuminated
layer. Indeed, the layer at $\tau_t\simeq 0.07$ is very much denser
than the layer immediately below it, and thus it is convectively
unstable.

The dotted curve in Figure (\ref{fig:thick}) was computed assuming
always high initial temperature ($k T_{\rm in} = 2$ keV). This curve
produces one sharp dip in temperature at $\tau_t\simeq 0.17$.  At this
point, only the cold solution exists. In the next spatial zone, due to
the larger line depth, there appears a solution with the intermediate
temperature, $kT\sim 100$ eV, which is found by XSTAR since the
initial guess for $T$ is high. However, there also exists a colder
solution, which is seen in the solid curve presented in the same
Figure. The latter curve was computed by providing XSTAR with the
initial temperature equal to that of the previous zone multiplied by
$1.5$. If the conductive heating was taken into account, it would force
the gas in the zone with the large jump in temperature to cool and so
it would lead to a solution close to the solid line. We show in the
Appendix that the conductive heating is always small except for
regions that are close to the temperature discontinuities. The latter
regions are always optically thin ($\Delta \tau_t \simlt 10^{-3}$ at
most, see Appendix), and hence they are unimportant from the point of
view of the radiation transfer and spectral reprocessing.

Out of these three methods of picking a solution, the one based on the
temperature of the previous zone is the most physical in the
sense that it has no sharp dips or spikes in temperature except for
the bump with the magnitude of about 2 in the region around
$\tau_t\sim 0.17 - 0.2$ (Similar features were earlier seen in the
work of Ko \& Kallman 1994). Therefore, we adopt this method
in the rest of the paper.

\section{Tests}\label{sect:simple}

\subsection{Setup}\label{sect:setups}
A systematic investigation of the reflected spectra as a function of
the accretion rate, $\dm$, and the magnitude of the X-ray flux, $\fx$,
is our eventual goal. Solving this problem consists of two steps: (1)
resolving the density and temperature profile of the illuminated
layer; (2) determination of how this profile influences the radiation
transfer in the illuminated layer, i.e., the reflected spectra.  Step
(1) is not necessarily difficult, but it depends quite sensitively on
the assumptions about the geometry of the X-ray emitting source (or
sources). For example, if X-rays are produced in magnetic flares,
then, because the geometry of the problem is no longer plane-parallel,
there is a possibility that a local wind is induced (see \S
\ref{sect:evaporation} on this). We clearly cannot treat this
situation within our current static approach. Another possibility is
that the accretion disk is not described by the standard model, such
that $H/R$ is different from the SS73 value, which then changes the
gravity parameter $A$. In other words, we have a large parameter space
to investigate, and such an investigation clearly goes beyond the
scope of this paper.

At the same time, for a fixed $\Gamma$, step (2) depends, to a large
degree, on two parameters only.  One of these is the Thomson depth of
the Compton-heated layer, $\tau_h$, and the other is the ratio of the
illuminating flux to the disk intrinsic flux, $\fx/F_d$, since this
ratio determines the actual value of the Compton temperature $T_c$.
Therefore, as a first attempt towards understanding of the effects of
the thermal instability on the reflected spectrum, we set up simple
tests that allow us to cover a range of values of $\tau_h$ and $T_c$
by varying one parameter at a time, thus isolating dependence of the
reprocessed spectra on that particular parameter.

We assume the following initial setup. We fix the accretion rate at a
value of $\dm = 10^{-3}$ (the disk is gas-dominated for such a small
accretion rate), the incident X-rays to be isotropic for simplicity,
the coronal dissipation parameter $f=0$ (no dissipation in the corona;
see Svensson \& Zdziarski 1984 for a definition), the black hole mass
to $M = 10^8 \msun$, and the incident X-ray flux (projected on the
disk surface) to $\fx = 10^{16}$ ergs s$^{-1}$ cm$^{-2}$. For these
values of parameters, $\fx/F_d \sim 1.6 \times 10^3$, and the gravity
parameter $A$ equal to $A_0 = 1.66 \times 10^{-3}$. To study the
effects of the thermal instability, we artificially vary the parameter
$A$ from this ``true'' value (other parameters of the problem will
later be varied as well). The rest of our methods, i.e., the radiation
transfer, the boundary conditions between the disk and the illuminated
layer are unchanged, that is the same as we will later use in fully
self-consistent calculations (\S \ref{sect:rd}).

\begin{figure*}[t]
\centerline{\psfig{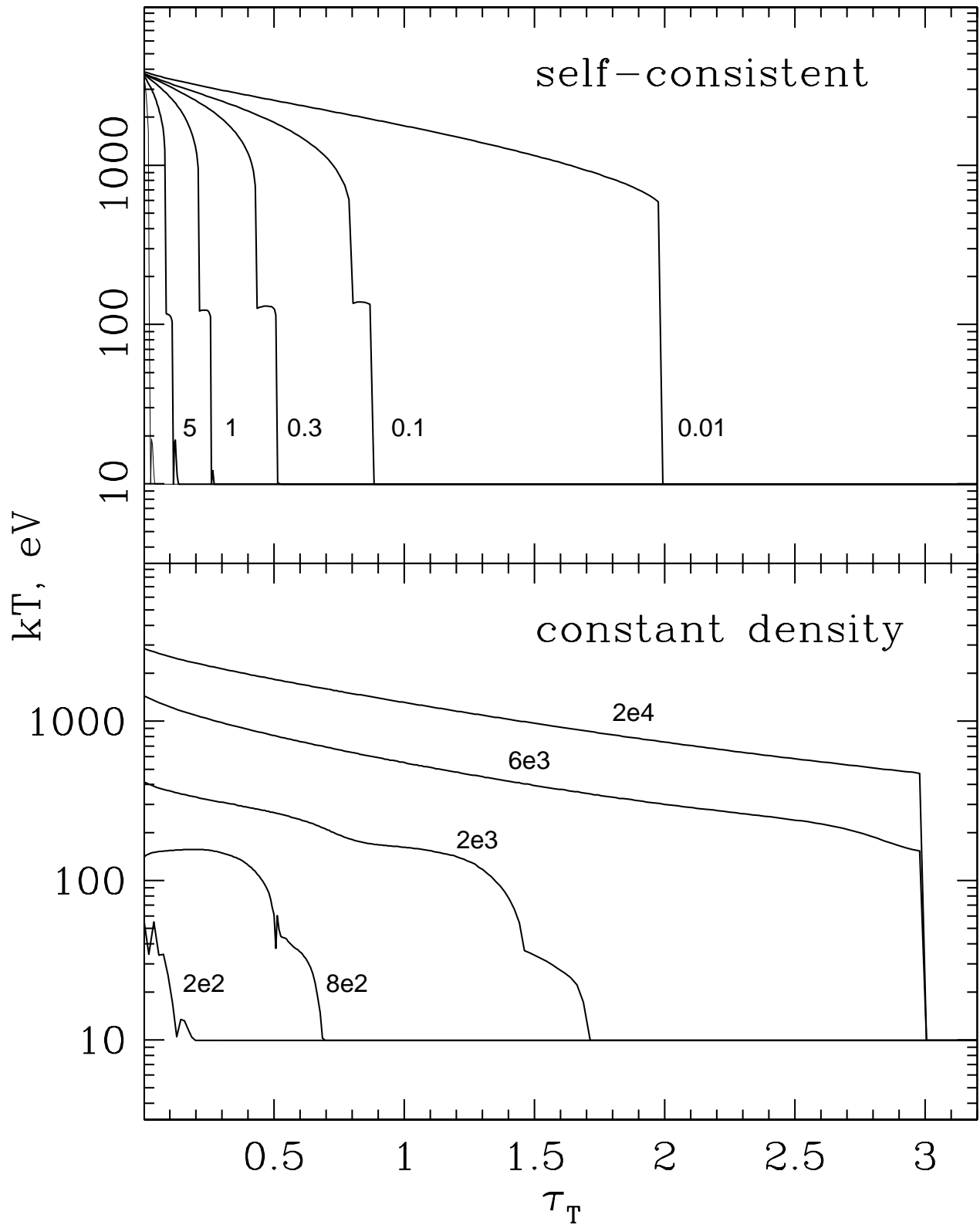}}
\caption{{\em Lower Panel:} Temperature of the illuminated gas
atmosphere as a function of the Thomson optical depth computed with
the constant density assumption. The parameter values are $\fx =
10^{16}$ erg cm$^{-2}$ s$^{-1}$ and $\Gamma = 1.9$ for all the tests. The
value of the density ionization parameter $\xi$ is shown next to the
corresponding curve. {\em Upper Panel:} Temperature of the illuminated
gas atmosphere computed with our self-consistent approach. The curves
differ by their respective values of the gravity parameter $A$, whose
value is shown to the right of the corresponding curve ($A=20$ for the
fine solid curve). Note that the self-consistent solution ``avoids''
the unstable regions of the S-curve, whereas the constant density
solutions unphysically cover the whole temperature range between the
effective and the Compton temperature.}
\label{fig:temper1}
\end{figure*}

\subsection{Hydrostatically stratified vs. fixed density atmospheres}
\label{sect:struct}

It is beneficial to start by comparing the results of our
self-consistent calculations with those obtained with the fixed
density approach, since the latter dominated the literature on the
topic during the past decade (e.g., Ross and Fabian 1993; \zycki et
al. 1994; Ross et al. 1997, 1999). The most important parameter in
photo-ionization calculations when the gas density is constant is the
density ionization parameter $\xi$ (eq. (\ref{xid}). In the case of
our calculations that include the hydrostatic balance, the ionization
structure of the illuminated layer depends most sensitively on the
``gravity parameter'' $A$. The upper panel of Figure
(\ref{fig:temper1}) shows temperature of the illuminated layer as a
function of the Thomson depth into the layer for several values of the
gravity parameter $A$, whereas the lower panel of the same figure
shows the gas temperature for different values of the density
ionization parameter $\xi$ (both the fixed density cases and the
self-consistent cases are computed with our code under equal
conditions and parameter values except for the gas density structure).

\begin{figure*}[t]
\centerline{\psfig{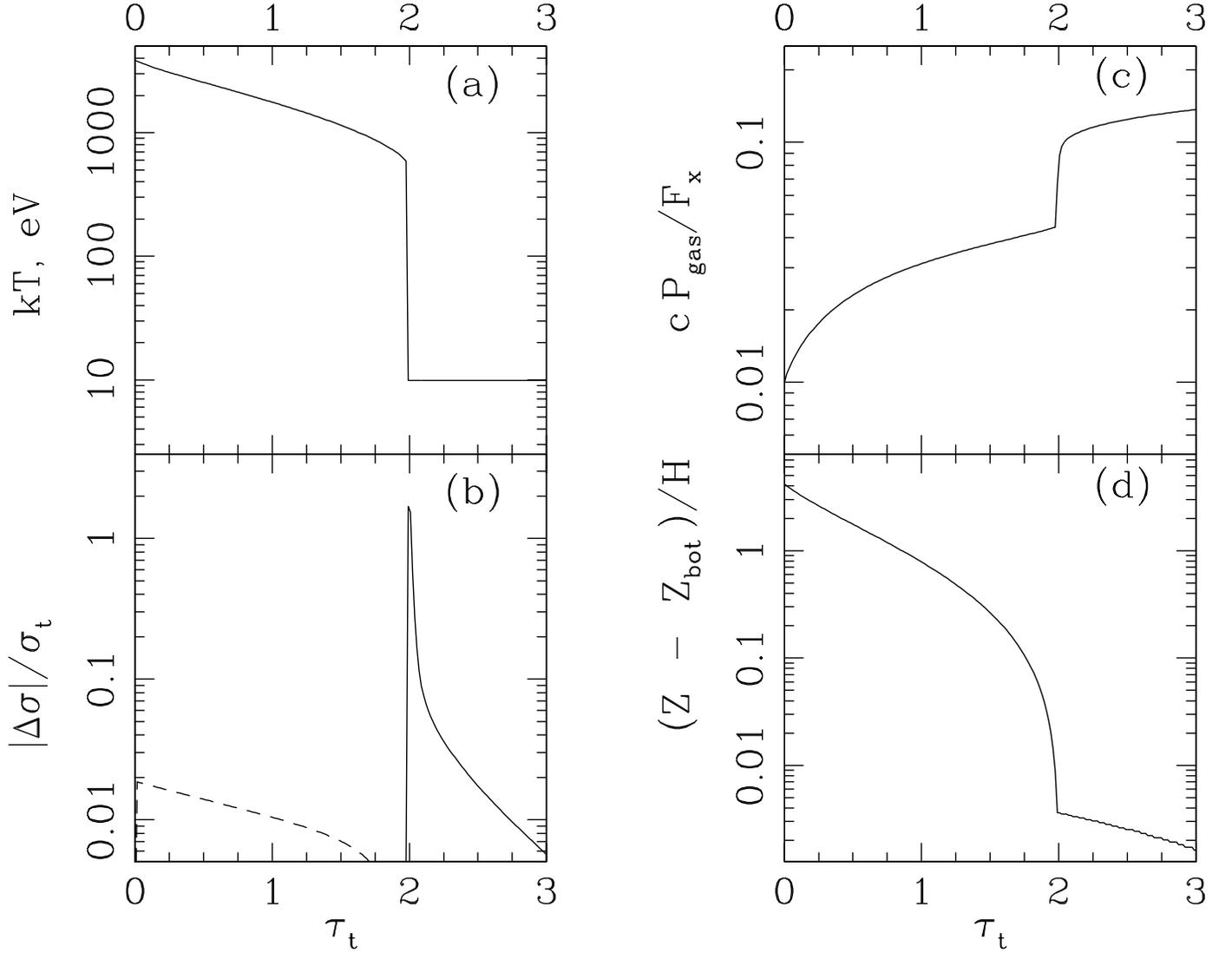}}
\caption{(a) Gas temperature, (b) radiation pressure force multiplier
(see eq. \ref{heq2}), (c) the gas pressure, and (d) the height above
the bottom of the layer ($z_b$) for the model with $A=0.01$. In panel
(b), positive values of $\Delta \sigma$ are depicted with a dashed
curve, whereas negative values are shown by the solid curve. Note that
the positive sign of $\Delta \sigma$ in the Compton-heated layer is
due to the relativistic roll-over in the Klein-Nishina cross section
for the incident hard X-rays.}
\label{fig:structure}
\end{figure*}

In short, the two sets of temperature curves have almost nothing in
common. The self-consistent solution avoids the regions in temperature
that correspond to a negative slope in the ``S-curve'' (cf. Fig.
\ref{fig:scurves}), i.e., the unstable regions, whereas they are
unphysically present in the fixed density cases
(Fig. \ref{fig:temper1}). To see that the fixed density solutions are
indeed unphysical, notice that since the gas density is constant, the
gas pressure is directly proportional to its temperature. Hence, on
the top of the illuminated atmosphere, instead of decreasing to zero,
the gas pressure exceeds that on the bottom of the layer by a factor
as large as a few hundred. The constant density calculations are out
of the pressure balance, and therefore will not exist in reality.

The temperatures that are allowed in the self-consistent solutions are
those with the positive slope in the ``S-curve''; the amount of
Thomson depth they occupy depends on the pressure balance.  Going from
the low illumination (high $A$) cases to the high illumination limit
($A\ll 1$), the amount of the hot material increases, as it should,
but the temperature of the Compton-heated layer does not change
considerably, except for the case when the hot layer becomes
Thomson thick. If one assumes that the hot material is ionized enough
to neglect line and recombination cooling, then it is possible to
analytically find the temperature, $T_t$, below which the Compton
heated solution ceases to exists (see \S IVb in KMT). The value of
$T_t$ is $T_t= 1/3 T_c$, i.e., a third of the Compton temperature at
the very top of the illuminated layer. The upper panel of Figure
(\ref{fig:temper1}) shows that this transition temperature is indeed
close to a third of the maximum temperature if $A\simgt 1$.  In the
opposite limit, the Compton-heated layer is optically thick and thus
the ionizing spectrum is actually different at the surface and at the
transition; the simple analytical theory would need to take this fact
into account in order to be more accurate.

In contrast, the fixed density calculations incorrectly predict that
not only the Thomson depth of the hot layer, but its temperature, too,
changes. This fact has obvious consequences for the reflection
component and the iron lines, as we will see below.

Figure (\ref{fig:structure}) shows the gas temperature, the
``radiation pressure force multiplier'' $\Delta\sigma $, the gas
pressure and the height above the bottom of the illuminated layer for
the self-consistent test with $A=0.01$. Fig. (\ref{fig:structure}d),
in particular, demonstrates that the density profile cannot be
assumed to be a Gaussian, because, in a rough manner of speaking,
there are two height scales -- one for the cold part of the
illuminated layer, and another for the Compton-heated part. 

Comparing the temperature profiles shown in the upper panel of
Fig. (\ref{fig:temper1}) with results of Ko \& Kallman (1994),
\rozanska and Czerny (1996), \rozanska (1999), one finds that the
intermediate stable solution occupies a smaller Thomson depth relative
to the Compton-heated solution than found by these authors. This small
Thomson depth is also somewhat unexpected because the part of the
S-curve between points (e) and (d) in Fig. (\ref{fig:scurves}) is not
``small'', i.e., it corresponds to a change in the gas pressure by a
factor of $\sim $ few. Panel (b) of Figure (\ref{fig:structure}) shows
that $|\Delta\sigma| \ll 1$ everywhere except for the region right
below the interface between the hot and the cold solutions, where the
X-ray radiation pressure force has a sharp peak. The direction of the
force is always down, i.e., in the same direction as the disk gravity,
because the soft X-ray opacity dominates over the opacity to the
quasi-thermal disk spectrum. Panel (c) of Figure (\ref{fig:structure})
shows that due to this X-ray pressure, the gas pressure increases
almost discontinuously once the temperature falls to the cold (or the
intermediate) stable branch values. Thus, the intermediate stable
branch of the S-curve can be passed very quickly and even ``skipped''
because of the additional radiation pressure.

Also note that the radiation pressure force due to the incident X-rays
minus that due to the reprocessed flux is small in the integrated
sense.  This allows us to approximately neglect this term in the
hydrostatic balance equation (\ref{heq2}). If the transition from the
hot to the cold solution occurs at $\Xi_t \sim$ few, then equation
(\ref{heq2}) tells us that the Thomson depth of the hot layer is
roughly
\begin{equation}
\tau_h\sim \left(\Xi_t A\right)^{-1}
\label{tauh}
\end{equation}
In accordance with this simple estimate, Fig. (\ref{fig:temper1})
shows that larger values of $A$ result in lower values of $\tau_h$,
although the dependence of gravity law on the height $z$ makes this
behavior more complex than equation (\ref{tauh}) predicts. A much
better analytical theory describing the temperature structure arising
from the self-consistent calculations can be obtained by using the
analytical theory of KMT and the equation for the hydrostatic balance;
we will present this theory in a future paper.

\begin{figure*}[t]
\centerline{\psfig{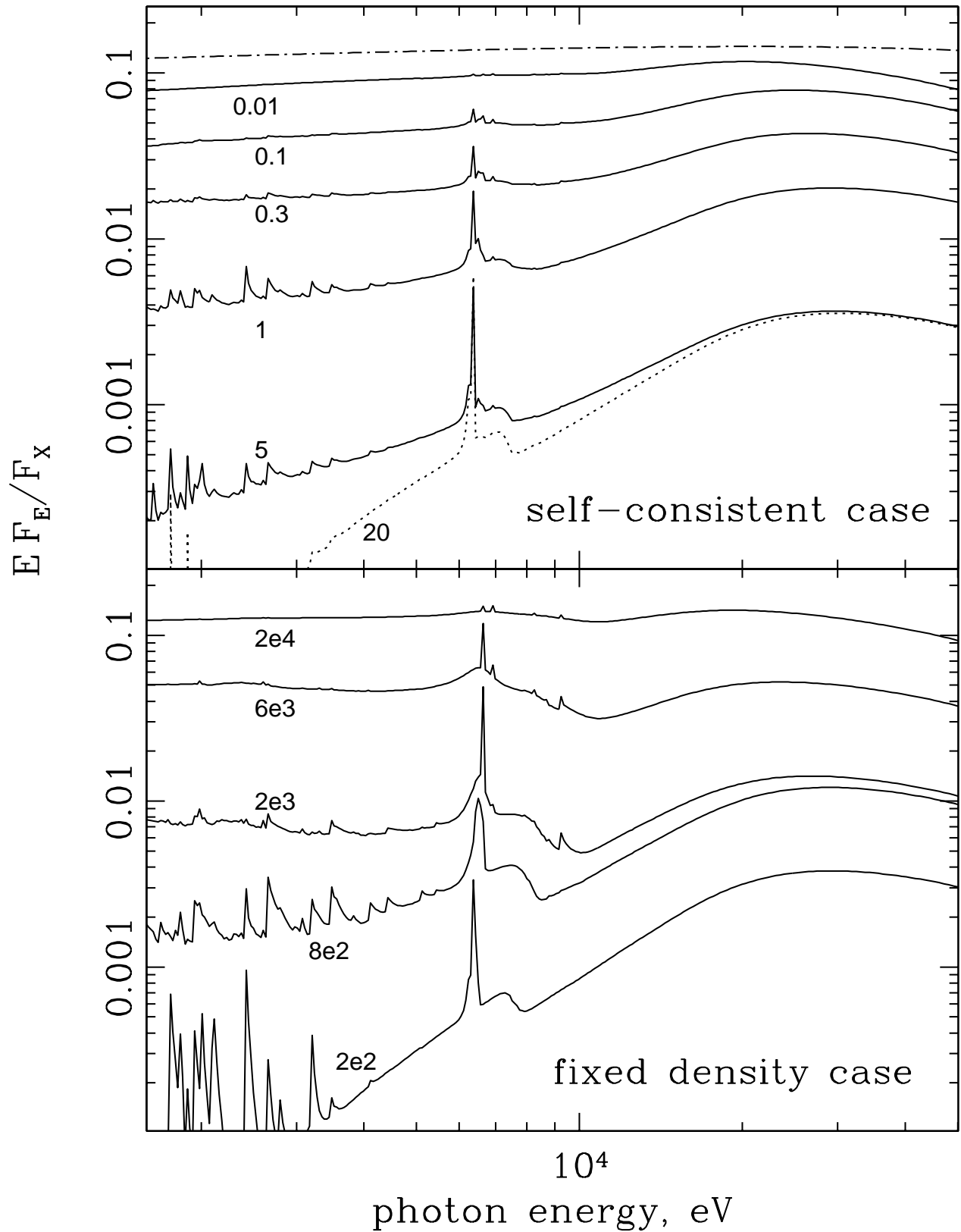}}
\caption{Reflected spectra for the tests presented in
Fig. (\ref{fig:temper1}). The incident ionizing spectrum is shown with
the dash-dotted curve. The curves are shifted by arbitrary factors for
clarity. Parameter values are printed next to their respective
curves. Note that evolution of the reprocessing features in the
constant density cases ({\em lower panel}) is distinctly different
from that for the self-consistent calculations ({\em upper panel}).}
\label{fig:spectra1}
\end{figure*}

\begin{figure*}[t]
\centerline{\psfig{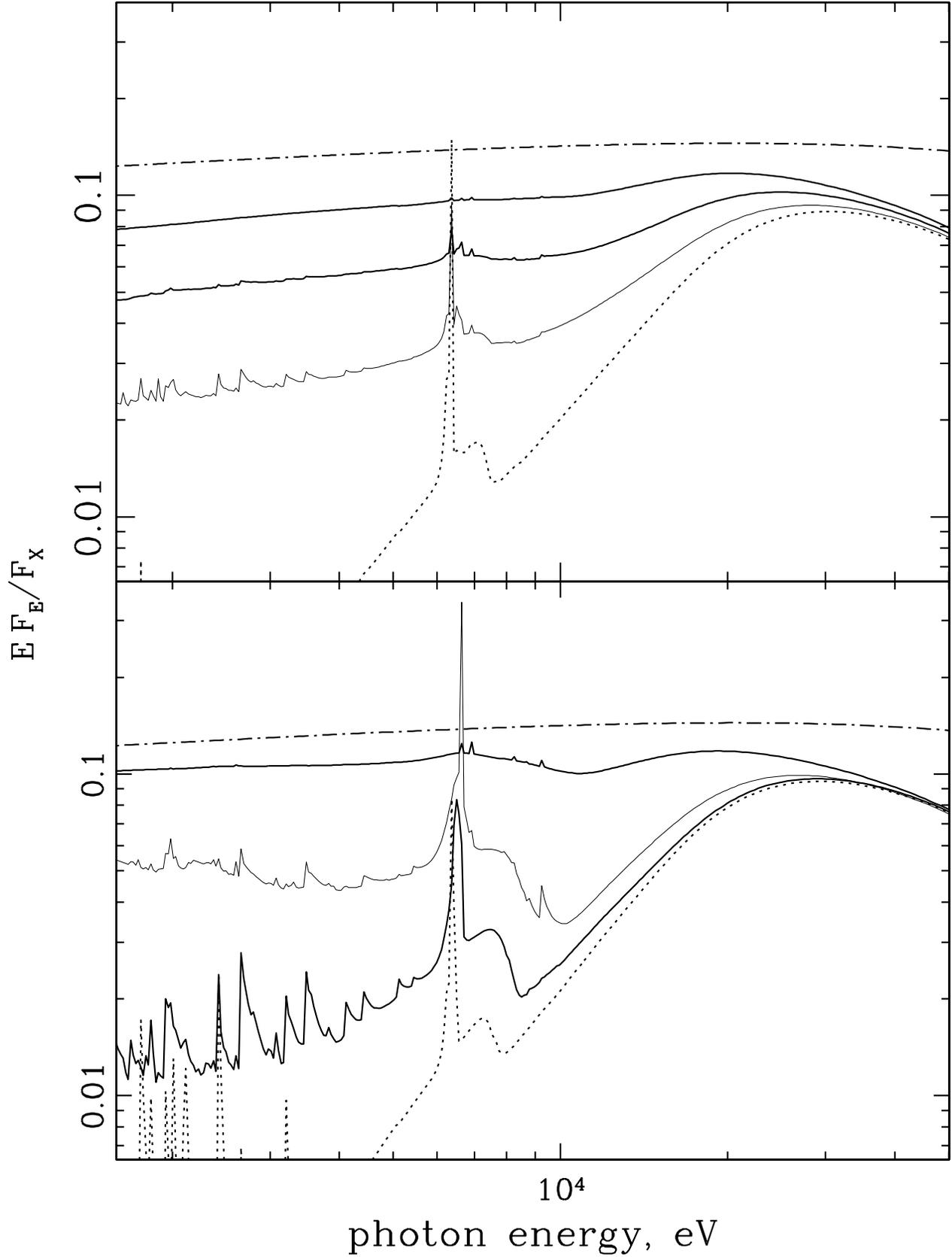}}
\caption{Selected curves from Fig. (\ref{fig:spectra1}) plotted with
the correct normalization. The incident ionizing spectrum is shown
with the dash-dotted curve. The parameter values for the curves are:
$A= 20$, 1, 0.1, 0.01 bottom to top for the {\em upper panel}, and $\xi
= 200$, 800, $2\times 10^3$ and $2\times 10^4$ bottom to top for the
{\em lower panel}.}
\label{fig:spectra2}
\end{figure*}

\subsection{Angle-Averaged Reflection Spectra}
\label{sect:aares}

\begin{figure*}[t]
\centerline{\psfig{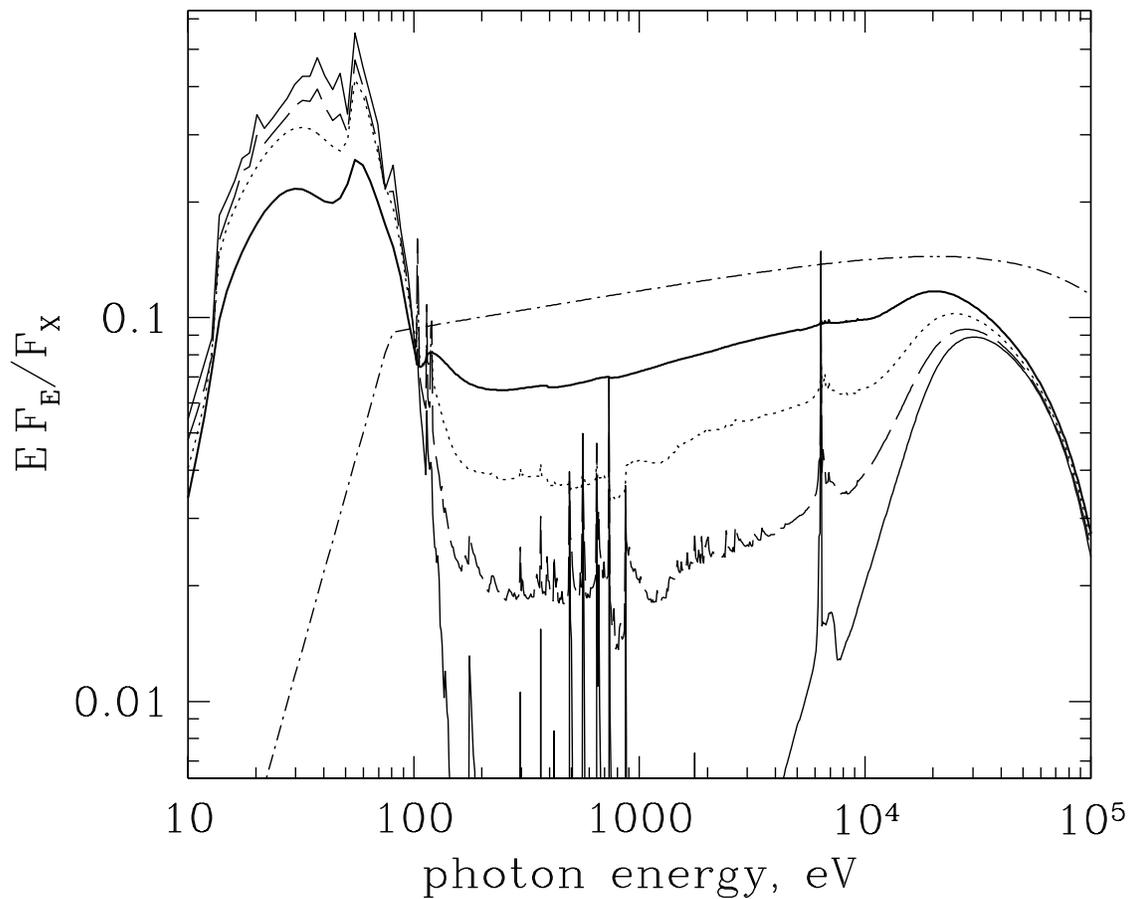}}
\caption{The angle-averaged reflection spectra for the tests shown in
the upper panel of Fig. \ref{fig:spectra1}. Note that the main effect
of the Comptonized layer is to fill in the gap (caused by
photoabsorption in the interval from $\sim 100$ eV to $\sim 10$ keV)
between the thermalized part of the reflected spectrum and the X-ray
reflection component.}
\label{fig:whole}
\end{figure*}

We will now discuss the reprocessed spectra that correspond to the
temperature profiles shown in Fig. (\ref{fig:temper1}). Figure
(\ref{fig:spectra1}) shows the reflected spectra for both the constant
density cases and our self-consistent calculations in the important
medium and hard X-ray energy range. These spectra are shifted with
respect to each other to allow easy visibility of the line. Also, in  
order to expose the true strength of the line and the reflected continuum, 
we re-plotted several of these curves with their correct normalization in
Fig. (\ref{fig:spectra2}). Finally, for completeness,
Fig. (\ref{fig:whole}) shows the same spectra as in
Fig. (\ref{fig:spectra2}) in the broader photon energy range.  It is
clear from Fig. (\ref{fig:whole}) that the reflected spectra in the
``thermal'' energy range (e.g., $\sim 10-100$ eV for the chosen X-ray
flux), are not accurately computed (as they never were in previous
works on the X-ray reflection). In order to improve the treatment of
the radiation field at these energies, one would need to take into
account stimulated emission and other many-body processes in the the
cold layers of the illuminated gas, which we plan to do in our future
work. Fortunately, spectra above $E\simgt 1$ keV should not be
affected by this.

The critical information needed to understand the results presented
herein is that for the relatively hard X-ray spectrum such as the one
with $\Gamma = 1.9$, most of the iron atoms in the Compton heated
layer turn out to be nearly completely ionized, and thus both line
emission and photo-absorption in this material are very weak. We can
define the following limiting cases to describe the behavior of the
reprocessing features.

\paragraph{High Illumination Limit.} When the
gravity parameter $A$ is small, i.e., $A\simlt 0.1$ no easily visible
reprocessing features result from the illuminated slabs. The physics
of this effect is rather transparent.  When the Compton-heated layer
is Thomson thick, the X-rays incident on the disk atmosphere are
reflected back before they can reach the cooler layers where the
material is less ionized and where they could be absorbed and
reprocessed into line or thermal radiation. Further, most of the
photons will be reflected back by the Thomson depth $\tau_t \sim 1-2$,
which then means that the Compton scattering changes the photon energy
only slightly unless the photon is mildly relativistic ($E\simgt 50$
keV). Therefore, it is to be expected that for photons of energy
smaller than the energy corresponding to the peak in the neutral
reflection component, i.e., $E\sim 30$ keV, the completely ionized
material represents a perfect mirror (only the photons' angle
changes after the scattering). Above $E\sim 30$ keV the photons are
downscattered in a way very similar to that in the neutral reflection,
so that the reflected spectrum above these energies is essentially
unchanged (see Fig. \ref{fig:spectra2}).

\paragraph{Low Illumination Limit.} The low illumination case can also
be called the ``strong gravity limit'', when $A\gg 1$. As Figure
(\ref{fig:temper1}) shows, the Thomson depth of the Compton layer is
much smaller than unity under these conditions, and hence its
existence is of no noticeable consequence for the reflected
spectra. The latter are basically the same as the reflection spectra
obtained in the classical studies of the neutral X-ray reflection as
far as the medium to hard X-ray photons are concerned, (e.g., Basko et
al. 1974; Lightman \& White 1988, Magdziarz \& Zdziarski 1995). The
iron line is rather narrow with centroid energy $\simeq 6.4$ keV, and
there is rather strong absorption edge above $\sim 7-8$ keV.

\paragraph{Moderate Illumination.} To a zeroth order
approximation, the intermediate cases, when illumination is neither
very strong or very weak, i.e., $0.1\simlt A\simlt 10$, represent a
combination of the low and the high illumination cases. The Thomson
depth of the Compton-heated layer is moderate ($\tau_h\simlt 1$), so
the X-rays propagating in the downward direction have a fair chance to
reach the cold layers. The X-rays incident on the latter have spectrum
similar to the incident one (except for the high energy end of the
spectrum), and therefore the spectrum that will be reflected off the
cold layers (i.e., the one propagating in the upward direction at the
interface between the Compton and the cold layers) will resemble the
cold non-ionized reflection. 

The Compton scattering of the ``reflection hump" in the hot layers
will not broaden it substantially.  However, the effects of Compton
scattering in the hot layers are quite pronounced for the energy range
from about 10 $\times k\teff$ to $\sim 10$ keV, because most photons
in this energy band would be absorbed in the standard cold
reprocessing (see, e.g., Fig. 5 in Magdziarz \& Zdziarski 1995, or the
lower solid curve in Fig. \ref{fig:whole}).  The larger the Thomson
depth of the hot layer, the larger is the number of photons of this
low energy ($< 10 $ keV) band that are Compton-reflected back within
the {\em hot upper layer}, reducing the apparent strength of the
``reflection hump". One should note that the influence of the
Compton-heated layer on the X-ray reprocessed spectra was already
discussed by Basko, Sunyaev \& Titarchuk (1974), but this issue was
not addressed at all in later works on X-ray reflection (and, in fact,
neither was the work of Basko et al.(1974)).

It is worthwhile to point out that the iron line centroid energy
remains at $E=6.4$ keV for all of our self-consistent calculations.
Similarly, the position of iron edge remains the same. These
reprocessing features only get weaker, and thus less visible as the
illumination increases ($A$ decreases). One can also note that there
is a certain amount of broadening of the line by Comptonization.  It
is possible that this broadening can actually be detected if it is
present in the data (Done 1999, private communications). In order to
accurately model line Comptonization, it is desirable to perform
higher resolution calculations using an exact Klein-Nishina scattering
kernel, which we plan to do in the future.

\subsection{Comparison to the constant density reflection spectra}
\label{sect:fixed_dens}

If the illuminating flux is low, then the material is non-ionized in
both self-consistent and constant density calculations, and its
temperature is close to the effective one. Thus, the spectra in the
low illumination limit discussed in \S \ref{sect:aares} should be
similar to those computed assuming constant gas density for $\xi\simlt
100$ or so (e.g., Lightman \& White 1988; \zycki et al. 1994;
Magdziarz \& Zdziarski 1995).  Likewise, if the illuminating flux is
very high, the gas temperature is close to the Compton one, and hence
irrespectively of whether the gas density is fixed or found from
hydrostatic equilibrium, the reflected spectra have very little iron
line or edge.  This is the limit of the Compton reflection, i.e., the
one without photo-absorption (cf. Basko et al. 1974, White et
al. 1988). The reflected spectrum has virtually the same power-law
index as the incident X-ray continuum, and only differs above
$E\simgt$ few tens of keV. Figures (\ref{fig:spectra1}) and
(\ref{fig:spectra2}) confirm these considerations.

However, the differences in the reprocessing features between the
self-consistent calculations with moderate values of the gravity
parameter, i.e., $0.1\simlt A\simlt 10$ and the constant density
calculations are truly remarkable (see Figs. \ref{fig:spectra1} and
\ref{fig:spectra2}). The former shows that the He-like iron line at
$\simeq 6.7$ and H-like line at $\simeq 6.9$ keV never dominate over
the ``cold'' iron line at $6.4$ keV, whereas the fixed density
calculations predict that at $\xi \simgt 800$ the iron line at 6.7 keV
will dominate. In addition, under the constant density assumption, the
Equivalent Width (EW) of the iron lines in the mildly ionized case can
be larger by a factor $\sim 2.5$ compared with the line's EW in a
neutral material (see Matt et al. 1993). On the contrary, our
calculations presented here show that solving for the pressure balance
leads to the opposite behavior of the EW. Going from low illumination
cases (large $A$) to higher illumination (lower $A$), one finds that
the strength of the iron lines rapidly decreases, and the absorption
edge gets less noticeable (cf. Fig. \ref{fig:spectra1}).

\subsection{Reflection at Different Viewing Angles}
\label{sect:vangles}

\begin{figure*}[t]
\centerline{\psfig{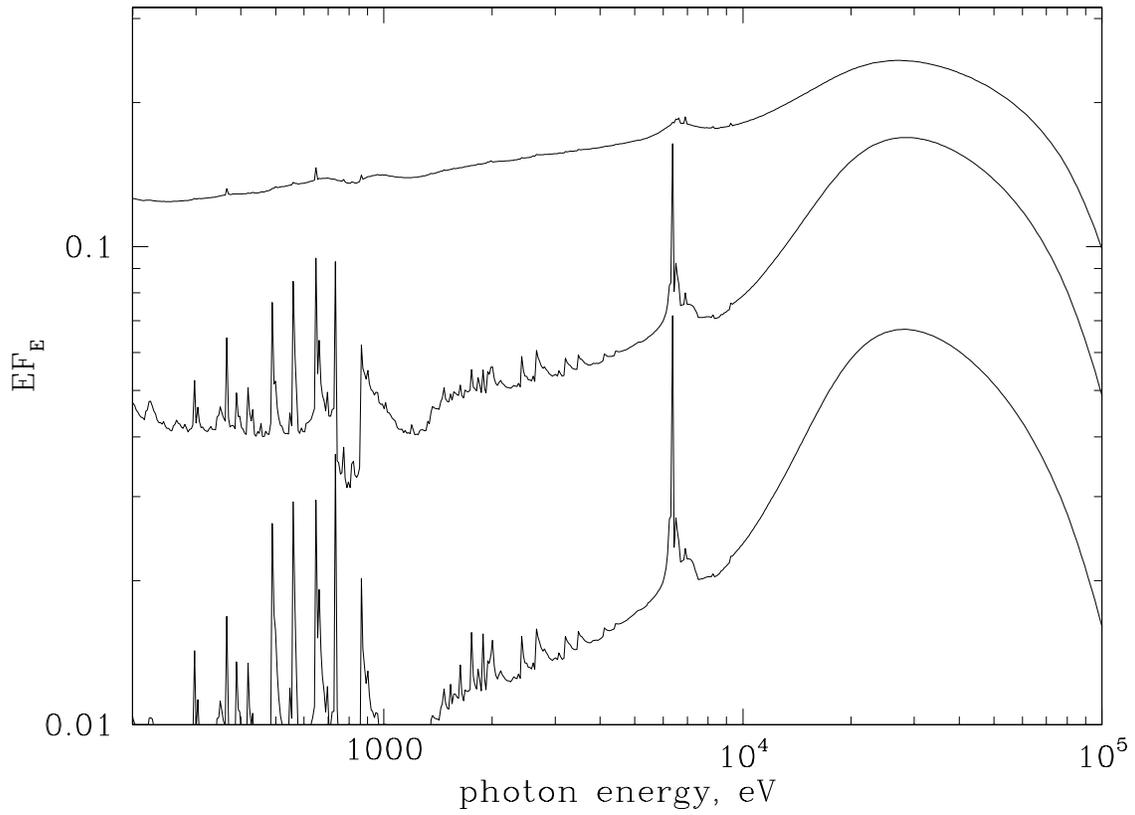}}
\caption{Reflected spectra (intensities) at different viewing angles
$i$ with respect to the normal to the disk for the self-consistent
case with $A=1$ shown in Figs. \ref{fig:temper1} \&
\ref{fig:spectra1}. The cosines of the angles are $\mu \equiv \cos i=
1/16$, $7/16$ and $15/16$ for upper, middle and the bottom curves,
respectively. For clarity of presentation, the middle curve is scaled
down by 1.2, whereas the bottom curve rescaled by 2.5.}
\label{fig:angle1}
\end{figure*}

Figure (\ref{fig:angle1}) shows the reflected spectra at three
different viewing angles for the test with $A=1$. It is seen that the
reprocessing features are strongest for the angle closest to the
normal, and that they almost disappear at large angles. The
explanation for this effect is rather straight forward: the Thomson
depth of the Compton heated layer, when viewed at an angle $i$ with
respect to the normal is $\tau_h(\mu) = \tau_h/\mu(i)$, where $\mu(i)
\equiv \cos i$, and $\tau_h$ is the Thomson depth of the Compton layer
as seen at angle $i=0$. Further, since one only receives emission from
material within one optical depth (roughly speaking), at large
inclination angles only the Compton heated layer contributes to the
reflected spectrum, and therefore no visible iron lines appear. This
effect should be important in modeling the iron lines from accretion
disks that are seen almost edge-on. Unless the optical depth of the
completely ionized layer is very close to zero, no iron line should be
observed in such systems (barring a more distant putative molecular
torus).

\subsection{Effects of Different X-ray Incidence Angles}
\label{sect:incident}

\begin{figure*}[h]
\centerline{\psfig{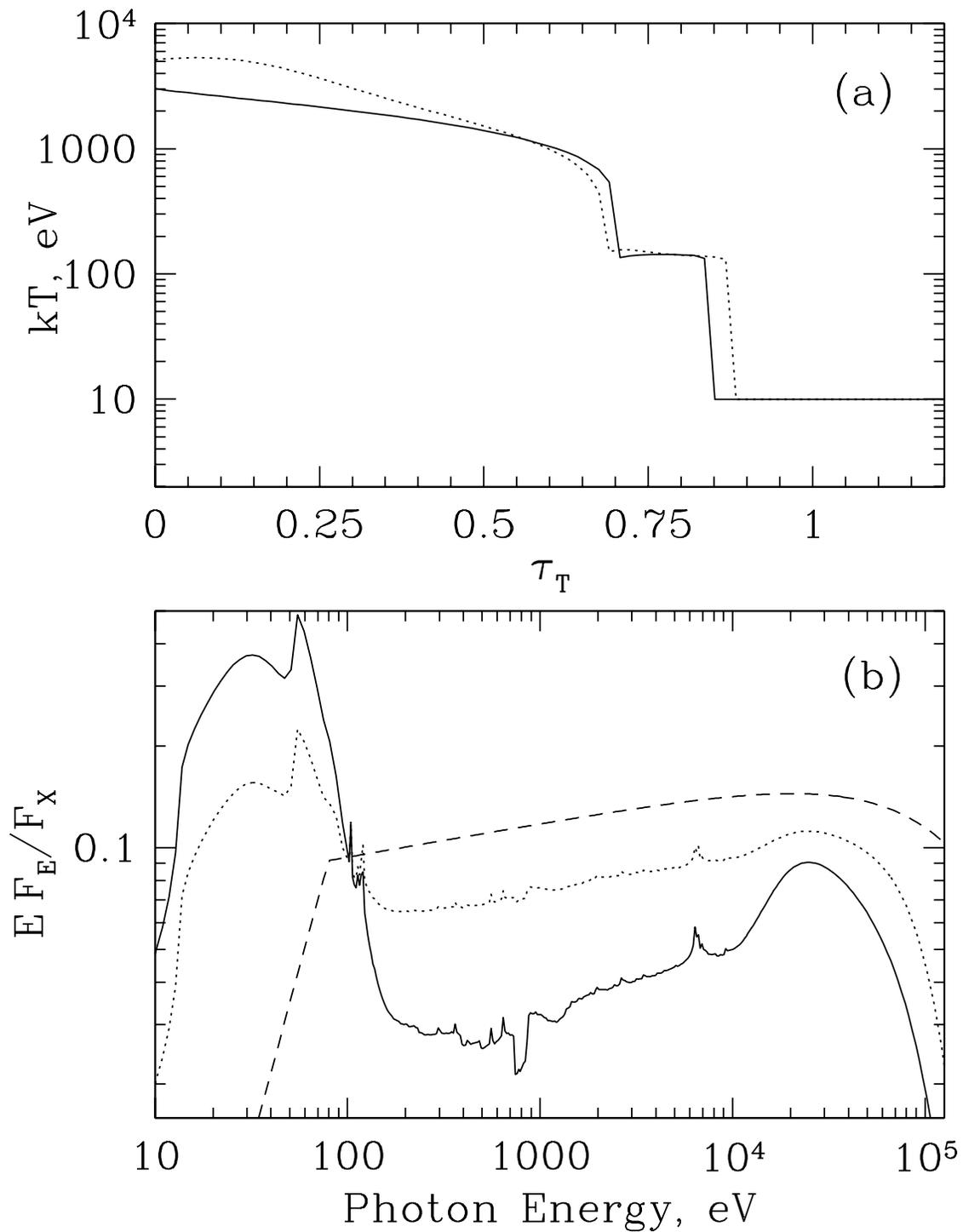}}
\caption{{\em Upper panel:} Temperature profile for the
self-consistent calculations with $A=0.1$ for different angles of the
X-ray incidence with respect to the normal to the disk.  The cosines
of the angles are $15/16$, and $1/16$ for the solid and dotted curves,
respectively. {\em Lower Panel:} Angle-averaged reflected spectra for
the two tests shown in the upper panel, together with the ionizing
spectrum depicted by the dashed curve.}
\label{fig:angle2}
\end{figure*}

Figure (\ref{fig:angle2}) shows the temperature profile and the angle
averaged spectra computed for $A= 0.1$, $\fx = 10^{16}$ erg s$^{-1}$ 
cm$^{-2}$, and $\dm = 10^{-3}$ for two different values of the X-ray
incident angle 
$\theta$\footnote{These spectra were computed with 400 photon energy
bins; i.e., with a lower resolution than most of the other
calculations presented in our paper.}. The solid curve corresponds to
$\cos \theta =15/16$ (i.e., almost normal incidence), whereas the dotted
curve corresponds to $\cos \theta =1/16$ (i.e., almost parallel to the disk
surface). It is interesting to note that the temperature profiles are
rather similar for moderate Thomson depths. We interpret this effect
as due to the fact that the X-rays are isotropized by the time
they penetrate to $\tau_t \sim 1$, so that the incident angular
distribution does not matter.

In the layer near the surface, however, the difference is quite
noticeable. In the case of the normal incidence, the gas temperature
is only $k T =  3$ keV on the very top of the layer, while it is
$k T = 5.2$ keV for the grazing angle of incidence. The difference is
easily explained by the fact that the Compton temperature is
approximately equal to
\begin{equation}
T_c \simeq {T_x J_x + \teff J_{\rm ref}\over J_x + J_{\rm ref}}\;,
\label{tcom}
\end{equation}
where $J_x$ and $J_{\rm ref}$ are the {\em angle integrated}
intensities of the X-ray and the reflected radiation (e.g., see
Begelman, McKee \& Shields 1983). We should now recall that the X-ray
flux $\fx$ is the flux projected on the direction normal to the
disk. Therefore, $\fx \propto \mu_i J_x$, whereas $F_{\rm ref} \propto
J_{\rm ref} \mu_{\rm ref}$. Note that $1/\sqrt{3}< \mu_{\rm ref} < 1$,
since the reflected radiation is clearly less isotropic than a truly
isotropic field that would have $\mu = 1/\sqrt{3}$ and yet less beamed
than a pencil beam escaping in the direction close to normal. Taking
into account that $F_{\rm ref}\equiv \fx + F_d \simeq \fx$, for X-rays
that are normally incident on the disk, $T_c \simeq T_x/(1 + \mu_{\rm
ref})$; for grazing angles of incidence with $\mu_i\ll 1$, $T_c \simeq
T_x/(1 + \mu_i/\mu_{\rm ref})\simeq T_x$. This is the reason why the
gas temperature is hotter for larger incidence angles. One can also
note that this region with $T\simeq T_x$ will not extend to very large
$\tau_t$, since at $\tau_t\sim \mu_i^{-1}$ most of the incident X-rays
will have been scattered at least once, which makes the X-radiation
field more nearly isotropic. This latter fact is clearly seen in
Figure (\ref{fig:angle2}a).

Turning to the difference in the reflected spectrum, Figure
(\ref{fig:angle2}b) demonstrates that the larger angles of incidence
produce more hard X-ray continuum and the reprocessing features are
weaker than in the case of nearly normal incidence. This effect is due
to a larger number of scatterings that the incident photons suffer
before they reach the cool line-producing layers in the former versus
the latter case; i.e. when the X-rays come in at a large
angle they are more likely to be reflected back before they can be
photo-absorbed in the cold material.

Finally, we should mention that for this particular calculation, we
turned off the X-ray pressure altogether, i.e., assumed $\Delta \sigma
\equiv 0$ in equations (\ref{heq1}) and (\ref{heq2}). As a result, the
intermediate stable state, i.e., the one with $kT\sim 150$ eV,
occupies a considerably larger range of the Thomson depth in
Fig. (\ref{fig:angle2}) than it did earlier for the tests shown in
Fig. (\ref{fig:temper1}), when we self-consistently computed $\Delta
\sigma$. This therefore confirms that it is indeed the X-ray pressure
that ``compresses'' the intermediate stable region (see \S
\ref{sect:struct}) compared to calculations of Ko \& Kallman (1994)
and \rozanska (1999).

\subsection{Different indices of the ionizing radiation}
\label{sect:indices}

\begin{figure*}[t]
\centerline{\psfig{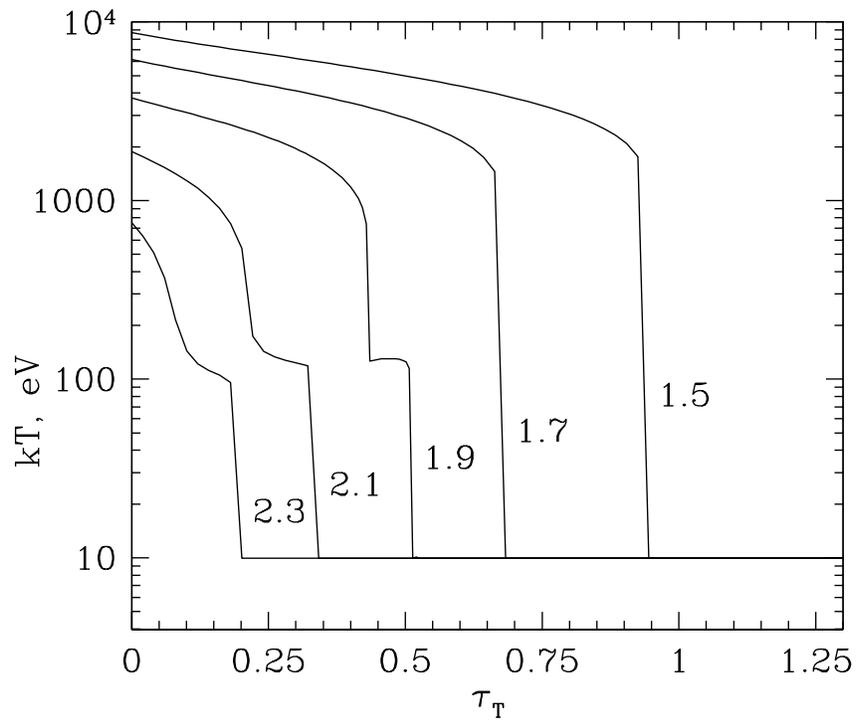}}
\caption{Temperature profile for self-consistently computed tests with
$A=0.3$, $\fx=10^{16}$ erg cm$^{-2}$ s$^{-1}$ and isotropic
illuminating flux for five different values of the spectral index
$\Gamma$ of the ionizing radiation.  Values of $\Gamma$ are shown to
the right of the respective curve.}
\label{fig:index}
\end{figure*}

Figure (\ref{fig:index}) shows how the temperature of the illuminated
layer varies with the hardness of the incident X-rays while all the
other parameters are held fixed. The fact that the maximum gas
temperature increases with decreasing $\Gamma$ is easily understood
because the Compton temperature of a harder spectrum is higher. In
addition, the harder the spectrum, the larger is the optical depth of
the Compton heated layer, because it takes more scatterings to bring
the average photon energy down such that the low temperature
equilibrium states will become possible. Also note that for a steep
spectrum, i.e., for $\Gamma\simgt 2.2$, the gas temperatures in the
range $\sim 200 - 1000$ eV become allowed (because for such a spectrum
the S-curve actually has a positive slope at these temperatures),
whereas they are forbidden for harder spectra.

\begin{figure*}[t]
\centerline{\psfig{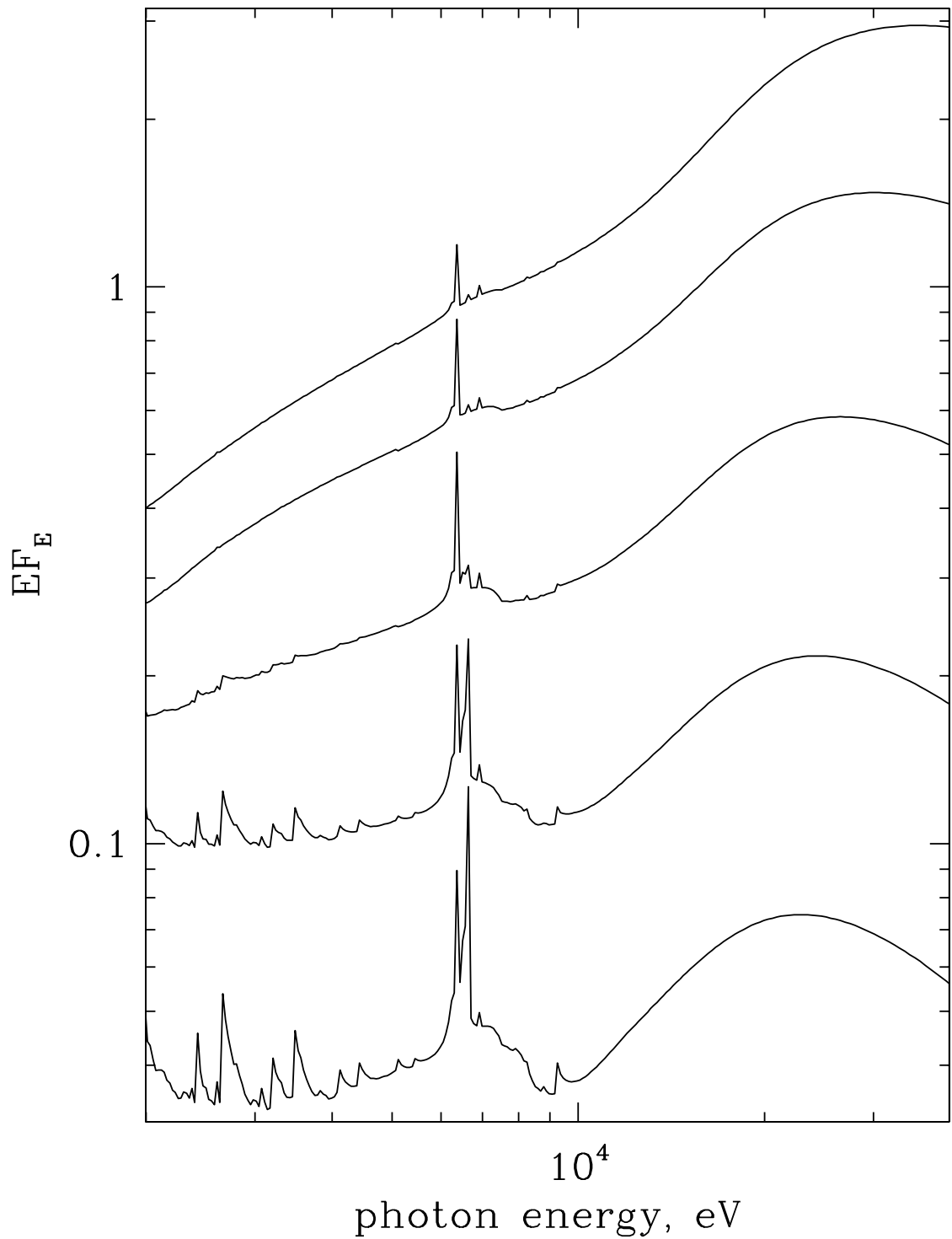}}
\caption{Reflected spectra for five different values of $\Gamma$ for
the tests presented in Fig. (\ref{fig:index}). The upper curve
corresponds to $\Gamma = 1.5$ and the lower one to $\Gamma=2.3$.
Curves are shifted with respect to one another to allow an easy
comparison.}
\label{fig:index2}
\end{figure*}

Because of these differences in the temperature profile, it comes as
no surprise that the reflected spectra are very different for hard and
soft spectra. In Figure (\ref{fig:index2}) we show the reprocessing
spectra corresponding to the five tests shown in
Fig. (\ref{fig:index}).  The hardest spectrum, i.e., $\Gamma=1.5$
shows very little iron line emission. It seems intuitively clear that
when one adds this spectrum to the illuminating X-ray spectrum, the
line and the reflection component will be barely discernible.  For a soft
X-ray spectrum, however, not only the 6.4 keV iron line is stronger,
but there is a strong 6.7 keV line as well.

\section{Radiation-Dominated Disks}\label{sect:rd}

\begin{figure*}[t]
\centerline{\psfig{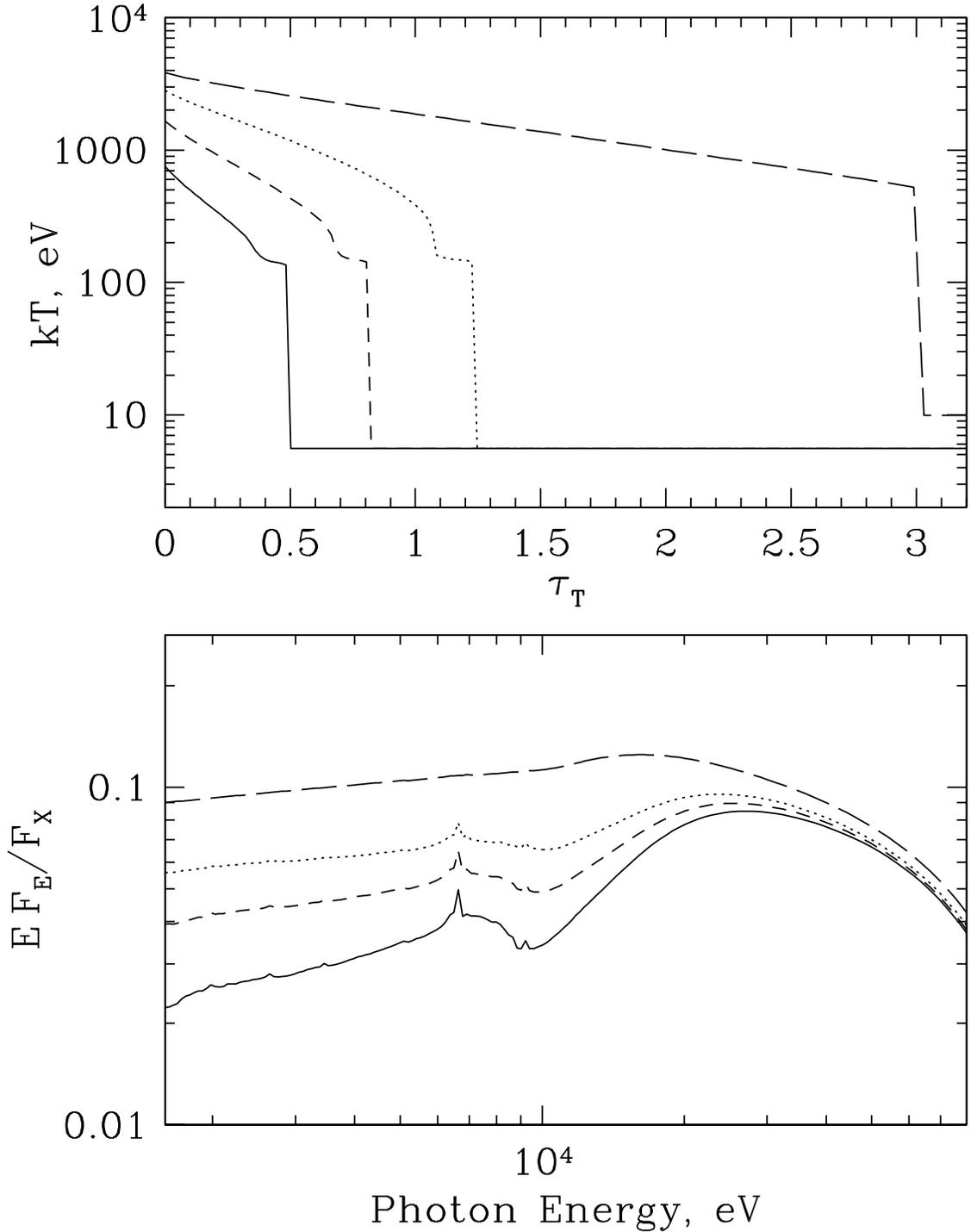}}
\caption{Temperature profiles (upper panel) and reflected spectra
(lower panel) for a radiation dominated disk accreting at $\dm = 0.1$
and $M_8 = 1$. The $A$-parameter is self-consistently found for a
given incident X-ray flux, which is equal to $10^{16}$ erg cm$^{-2}$ s$^{-1}$
for the long-dashed curve, and $10^{15}$ erg cm$^{-2}$ s$^{-1}$ for the rest
of the curves. The dotted curve is computed for the self-consistent
value of the disk intrinsic flux $F_d$ (i.e., $F_d/\fx = 0.63$ for
this case), whereas it is 4 and 12 times this value for the
short-dashed and solid curves, respectively.}
\label{fig:rd}
\end{figure*}

We now show several tests for a radiation-dominated accretion disk
where the parameter $A$ was equal to its nominal value as defined by
equation (\ref{adef}). Figure (\ref{fig:rd}) shows the gas temperature
profiles and the reprocessed spectra for two values of the incident
X-ray flux ($10^{16}$ erg cm$^{-2}$ s$^{-1}$ for the long-dashed curve, and
$10^{15}$ erg cm$^{-2}$ s$^{-1}$ for the rest of the curves)\footnote{These
tests are conducted with 400 energy points as well.}.

Let us first concentrate on the long-dashed and the dotted
curves. These differ only by the absolute magnitude of the
illuminating flux. The self-consistent value of $A$ for these tests is
$A = 6.34\times 10^{-2}$ and $0.634$ for the long-dashed and dotted
curves, respectively. If we now compare these curves with the sequence
of curves shown in the upper panel of Fig. (\ref{fig:temper1}), then
we can observe that the radiation-dominated (RD) illuminated layers
have much thicker Compton-heated regions than Fig. (\ref{fig:temper1})
would predict for similar values of $A$. The difference is due to the 
fact that the disk radiation force is strong even above $z=H$ for a RD
disk: the radiation force due to the disk intrinsic emission cancels
most of the gravitational force in the vicinity of the disk surface
(see equation \ref{heq2} and \S \ref{sect:pbc}). It is then convenient
to introduce a ``modified gravity parameter'' $A_m$ defined as
\begin{equation}
A_m\equiv A - {F_d\over F_x}\;,
\label{amoddef}
\end{equation}
which then allows one to rewrite the equation for the hydrostatic
balance for the illuminated layers above  the RD disks as
\begin{equation}
{\partial {\cal P}\over\partial \tau_H} = \,\left[A_m + A\,{z-H\over
H} \, - \, {\Delta\sigma \over \sigma_t} \right]\;.
\label{heq3}
\end{equation}
In the simulations presented in figure (\ref{fig:rd}), the values of
$A_m$ were $\simeq 2\times 10^{-3}$ and $2\times 10^{-4}$, for the
dotted and long-dashed curves, respectively. In general, both terms
($A_m$ and $A$) are important in establishing pressure
balance, so that the classification of the reflected spectra by only
one parameter $A$ is not as straight forward as it was for a
gas-dominated disk (\S \ref{sect:simple}). Nevertheless, the behavior
of the reflected spectra with $A$ is qualitatively the same.

To investigate the importance of cooling due to the disk intrinsic
flux, we set up two additional calculations. We kept $A_m$ and $A$ the
same as for the dotted curve in Figure (\ref{fig:rd}), but increased
the intrinsic disk flux by factors 4 and 12 for the short-dashed and
solid curves, respectively. This test demonstrates that the usually
unstable temperature region between $\sim 200$ eV and 1 keV can become
thermally stable not only due to a soft X-ray incident spectrum (see
\S \ref{sect:indices}), but also due to a large UV flux. As one can
deduce from equation (\ref{tcom}), the large intrinsic disk flux makes
the maximum Compton temperature equal to $T_c \simeq T_x \fx/F_d\ll T_x$.
This means that the ionized layer may have a relatively low
temperature. When this temperature is below $\sim 1$ keV, the
reflected spectrum will contain features characteristic of ``highly
ionized'' reprocessing (if the layer is somewhat optically thick),
even though the ionizing spectrum is hard, in contrast to the results
obtained for $\fx/F_d\gg 1$ (cf. \S \ref{sect:indices}).

\section{Gas evaporation and other effects}\label{sect:evaporation}

So far we have been dealing with the problem of X-ray illumination in
the context of a plane-parallel geometry where the X-rays come from
infinity at a given incidence angle. Some of the possible geometries
for the X-ray source and the accretion flow are not very far from this
setup, while others may require solving a dynamical rather than a
static problem. For example, if the X-ray source is located at some
height above the black hole (``lamp post model''), as for example, in
\cite{rb97}, then the reprocessed spectra can be obtained by our
methods if one also takes into account the variation in the ionizing flux
$\fx$ and the incident angle with radius. One would then calculate the
reflected X-ray spectra for a grid of radii and then integrate over
the disk surface.

However, a difficulty that cannot be resolved by our methods appears
in the context of an accretion disk model where X-rays come from
localized magnetic flares (e.g., Galeev et al. 1979; Field \& Rogers
1993; Haardt et al. 1994, Nayakshin 1998c; Nayakshin \& Dove
1999). This difficulty is that the problem becomes not static because
evaporation of the material into a {\em wind} may become important.

Let us elaborate a little on this issue. As we mentioned, the thermal
gas pressure of the Compton-heated layer cannot drive a wind out to
great distances because the Compton temperature $T_x\simlt 10^8$ K is
much below the virial temperature ($T_{\rm vir}$ of the gas in the
inner disk $T_{\rm vir} = 2.7 \times 10^{12} r^{-1}$ K (see also
Begelman, McKee \& Shields 1983). Similarly, the X-radiation for a
sub-Eddington accretion rates (which we implicitly assumed here)
cannot drive a global wind. Finally, line-driven winds (e.g., Proga
1999; Proga, Stone \& Drew 1998) also cannot remove the Compton heated
material since it is highly ionized and the line pressure force is
basically zero. What the X-radiation pressure and the thermal pressure
can do is to move the Compton-heated gas upwards or side-ways on
distances of the order of few disk height scales $H$ (one can check
that moving the gas farther is counter-acted by the gravity and hence
cannot be done). As far as we can see, this fact is of no consequence
in the case of a single central X-ray source, whether it is located
above the black hole or it is in the disk plane (e.g., Fig. 2 in Dove
et al. 1997), since the typical dimensions of the X-ray illuminated
region are $\sim R\gg H$.

In contrast, moving the hot gas away from a magnetic flare, the source
of the illuminating X-rays, on the distance of order few $H$ may be
quite significant for the disks with flares, because most of the X-ray
reprocessing will be done within this distance from the flare (see
Fig. 1 in Nayakshin \& Dove 1999 for an illustration). Once far enough
from the flare, the gas will cool and settle down on the disk without
much observational consequences. At the same time, the removal of the
Compton-heated gas away from the flares reduces the value of the
Thomson depth below the values calculated in this paper. Therefore,
 value of the gravity parameter $A$ alone does not define the
reflected spectra in the case of a magnetic flare, and future
calculations, including gas dynamics are needed.

Fortunately, the reflected spectra depend most sensitively on the
Thomson depth, $\tau_h$, of the Compton layer between the source of
X-rays and the cold material, irrespective of whether a particular
value of $\tau_h$ is obtained in a static situation or in the ''local
wind''. The wind velocity is comparable but somewhat larger than the
thermal sound speed of the Compton-heated gas (see \S III of McCray \&
Hattchet 1975 in the context of a stellar X-ray induced wind), which
is only $c_{\rm hot}\simeq 3\times 10^{-3}$ of the speed of
light. Because the gas in the wind will probably be Compton-heated, so
that line opacities and emission are not important there, we do not
expect any significant effects due the gas bulk motions on the
radiative transfer.

Accordingly, the X-ray reprocessing spectra should be very similar to
the ones we obtained in this paper, covering the same low, moderate
and high illumination limits, but the relation of these limits to the
accretion disk and flare parameters (e.g., $\dm$, $\fx$, size of the
flare, etc.) is yet to be clarified.

\section{Summary}\label{sect:summary}

We now summarize the important points about X-ray reprocessing and
iron line emission emerging from the calculations presented in this
paper. We also discuss several observationally important issues where
our methods should find useful applications.

{\bf 1.} The thermal ionization instability plays a crucial role in
determining the density and temperature structure of the illuminated
gas. Its effects can be taken into account by a proper treatment of
the hydrostatic balance or dynamics of winds if the latter are
important.  Because of a (nearly) discontinuous behavior of the gas
density and temperature caused by the thermal instability, one may not
assume the illuminated gas to have either a constant or a Gaussian
density profile.

For {\bf hard X-ray spectra} with $\Gamma\simgt 2$, and large X-ray
flux (i.e., $\fx/F_d\gg 1$), the thermal instability ``forbids'' the
illuminated layers from attaining temperatures in the range from 
$\sim 200$ eV to $\sim T_c/3$ typically
$\sim$ few keV. Temperatures in the intermediate range $\sim 80 -
200$ eV are allowed, but due to the X-ray radiation pressure effects,
the layers with these intermediate temperatures occupy a small Thomson
depth and thus contribute little to the reflected spectra. The major
contributor to the reprocessing features (lines, edges and the
characteristic reflection bump itself) is the cool layer with
temperature close to the effective temperature for the given X-ray and
disk fluxes. The overall structure of the illuminated layer is then
approximately two-phase one (neglecting the intermediate temperature
layer): (i) the Compton-heated ``skin'' on the top of the disk, and
(ii) cold, dense, and thus typically weakly ionized layers below the
hot skin.

The iron is completely ionized in the Compton-heated material (for
hard spectra), and thus the only important radiative processes there
are Compton scattering and bremsstrahlung emission. As a result, the
existence of the hot material always reduces the strength of the
reprocessing features, because the incident X-rays may scatter back
(and out of the disk atmosphere) before they reach the cold layers
where atomic processes could imprint the characteristic reprocessing
features. The Thomson depth, $\tau_h$, of the Compton-heated
illuminated layer is controlled by the gravity parameter $A$, 
whose values distinguish the following characteristic cases:
\medskip

\noindent$\bullet$ Low illumination, i.e., high values of $A$ $(A\gg 1)$,
which result in weakly ionized matter and lead to $\tau_h\ll 1$ and hence
to iron lines, edges and continuum reflection characteristic of neutral
material.
\medskip

\noindent$\bullet$ Moderate illumination, which corresponds to values of 
$A$ in the range $0.1 \simlt A\simlt 10$, and produce moderately thick
Thomson-heated layers. As far as the reflected continuum is concerned,
the reprocessed spectrum represents a combination of that for the high
and low illuminated cases discussed above. This may not be true for
the iron lines, though, because Compton scattering of the line photons
in the hot layer can be very effective in removing these from the line
into the continuum (Done 1999, private communications). Further work
is needed in order to understand how this moderate illumination limit
will appear when observed by a particular instrument. We believe that
because the Fe atomic features are created in the cold nearly ``neutral''
material, and yet their normalization is lower than that of 
standard neutral reflection, some narrow band X-ray telescopes may
confuse the mildly illuminated reflector covering a full $2\pi$ solid
angle with a {\em non-ionized} reflector covering a fraction of
$2\pi$.

\medskip

\noindent$\bullet$ High illumination ($A < 0.1$). X-radiation ionizes
a substantial amount of material, so that the Compton-heated layer is
Thomson-thick.  Very little of either of the 6.4, 6.7 or 6.9 keV iron
lines are created in this layer because few X-ray make it through to
the cold disk material. Thus, the reprocessing features, except for
the roll-over at few tens keV due to the Compton down-scattering are
wiped out of the reflection spectrum. The latter is a power-law with a
similar index as the incident one for $E\simlt 30$ keV, and can be
undetectable in the lower energy data.

\medskip
\noindent$\bullet$ The evolution of the reflection component and the
iron lines from the weak illumination limit to the strong illumination
limit is monotonic and in no point does the spectrum exhibit
observational signatures of ``highly ionized'' matter (as do the 
spectra of the constant density studies), because
the Compton layer is completely ionized and the line-creating material
is very cold, effectively neutral. This is to be contrasted with
the predictions of constant density models, where the EW of the
line and its centroid energy increases with ionization parameter $\xi$
(e.g., Ross \& Fabian 1993; Matt et al. 1993; \zycki et al. 1994;
Ross et al. 1999). 

\medskip{\bf 2.} Ionized iron lines and strong absorption edges can
nevertheless be produced under the following conditions:

\medskip
\noindent$\bullet$ If $A\ll 1$, and the incident X-ray spectrum is
steep, as in soft states of GBHCs, i.e., $\Gamma\simgt 2$, then the
Compton temperature even on the top of the reflecting layer can be
lower than $\sim 1 $ keV, which can then lead to the appearance of
$6.7$ and $6.9$ keV iron lines.

\medskip
\noindent$\bullet$ If $A\ll 1$, and the incident X-ray spectrum is
hard, $\Gamma\simlt 2$, but the disk intrinsic flux exceeds the X-ray
illuminating flux, i.e., $F_d \gg \fx$. This again lowers the Compton
temperature of the hot layer and hence could yield ionized lines and
strong absorption edges.

\medskip{\bf 3.} Reflected spectra viewed at larger angles relative to
the disk normal always contain less reprocessing features (e.g., lines
and edges) than those viewed face-on.

\medskip

Our findings are of interest for the modeling and the interpretation of 
observations from a number of astrophysical sources:

Recently, an X-ray ``Baldwin'' effect was discovered by Iwasawa \&
Taniguchi (1993) in the{\rm Ginga} data and with a much better
statistics by Nandra et al. (1997b) in the {\em ASCA} data: The 6.4
keV Fe line and the reflection component in the Seyfert 1 X-ray
spectra, which typically have $2-10$ keV luminosities $L_{2-10}
\simlt$ a few $\times 10^{44}$ ergs s$^{-1}$ can be well fit with
reprocessing arising in low-ionization parameter material (Nandra et
al. 1997a). Higher luminosity sources exhibit a {\em monotonic
decrease} of the Fe line equivalent width (EW) (Nandra et al. 1997b).
Above $L_{2-10}\simeq 10^{45}$ ergs s$^{-1}$, the EW of the line drops
abruptly and this line becomes undetectable for $L_{2-10}\sim 10^{46}$
ergs s$^{-1}$, both in radio loud and radio quiet quasars. These
observations are consistent with the general trend of our calculations
which indicate that upon increase of the X-ray flux, {\em the X-ray
reflecting material converts from ``cold'' to completely ionized},
with a concurrent decrease in the strength of the Fe 6.4 line, for the
reasons discussed above.  By contrast, models assuming a constant
density for the illuminated gas predict that such an increase in the
X-ray flux should give rise to the presence of intermediate states
with even higher EW of He- and H-like iron lines and a deep broad iron
edge (see Fig. \ref{fig:spectra1}, and also Fig. 1 in Matt et
al. 1993; Fabian 1998), which are not seen in the data. Our results
show that, when computed self-consistently, i.e., using hydrostatic
pressure balance, the line EW and the depth of the edge monotonically
decrease in accordance with the observations. In future work we plan
to explore in detail the interpretation of this effect in the context
of general, global AGN models.

Another  application of our code is the modeling of the reprocessing
features in transient BH sources, because these cover a wide range of
accretion rates $10^{-4} < \dm \leq 1$ while they decline from the
maximum luminosity. In studying the X-ray reprocessing features and their
relativistic smearing, one can potentially constrain such parameters
of the accretion flow as the inner radius at which the disk terminates.
This kind of modeling was in fact performed by \zycki, Done
\& Smith (1997), who studied reprocessing features in GS 2023+338
during its decline \zycki et al. (1998) who did the same for the
Nova Muscae 1991 observations; and by Done \& \zycki (1999) for
Cyg~X-1. Given the results of this paper, it seems rather likely that
the X-ray reprocessing calculations that include hydrostatic balance
can make a noticeable difference in terms of how close to the black
hole the ``cold'' matter is in these sources, and what is the
evolution of the flow geometry during the luminosity decline.

Narrow Line Seyfert Galaxies (NLSG), that typically have steep X-ray
spectra, may provide an additional opportunity to test accretion
flows. As we found, steep X-ray spectra may lead to a sufficiently low 
Compton temperature, i.e., $kT_c\simlt 1$ keV, which then allows some
of the highest ionization stages of iron ion to be abundant enough to
imprint ``highly ionized'' signatures on the X-ray spectra. In other
words, due to steeper X-ray spectra, the hot Compton layer may now
become visible (it is ``invisible'' for other objects with harder
spectra because it does not produce any sharp line or absorption
features), thus yielding interesting constraints on accretion flow
geometry. Some of the observed NLSG indeed reveal unusual ionization
states (see, e.g., Comastri et al. 1998 and Bautista \& Titarchuk
1999).

\section{Conclusions}\label{sect:conclusions}

In conclusion, we presented above a physically consistent way of
computing the reprocessing features for any accretion disk
geometry/model that does not produce X-ray driven winds. An important
point to note is that the reflected spectra depend mostly on the
gravity parameter $A$, which is determined by the illuminating flux,
and the local gravity (or the ratio $H/R$). The latter is independent
of $\alpha$-parameter -- the usual Shakura-Sunyaev viscosity parameter
whose value is notoriously uncertain (see, e.g., Nayakshin, Rappaport
\& Melia 1999). Therefore, at least for a range of conditions (to be
specified in a future paper), the resulting reprocessed X-ray spectra
are independent of $\alpha$ and thus can be unambiguously calculated
for a given accretion disk model. Therefore, we believe that the
methods presented in this paper may be used as a meaningful and
sensitive test of theoretical predictions versus observations of the
iron lines, edges and the reprocessing features in general. With
advent of the new X-ray telescopes (Chandra, Astro-E and XMM), such
tests may bring us one step closer to establishing which (if any) of
the current accretion disk models is correct.

\bigskip

SN acknowledges support by an NAS/NRC Associateship during the course of 
this work. The authors thank Manuel Bautista, Chris Done, Julian Krolik, 
Daniel Proga, Lev Titarchuk and Piotr \zycki for many useful discussions. 

\appendix

\section{On The Importance of Thermal Heat Conduction}\label{sect:cond}

We can estimate the maximum Thomson depth, $\Delta \tau_m$, of the
region where the heat conduction is important by the following
considerations. Suppose that $\Delta z$ is the length scale over which
the transition from the hot ($T\sim 10^8$) solution to the cold one
($T\sim 10^5$ ) occurs. The maximum heat flux due to conduction is the
saturated conductive flux,
\begin{equation}
F_{\rm sat} \sim c_h P\;,
\label{satf}
\end{equation}
where $c_h$ is the isothermal sound speed at the hot (Compton)
solution, and $P$ is the gas pressure (see Cowie \& McKee 1977, and
also Max et al. 1980). The X-ray heating per unit volume can be
estimated as $Q^+_x\sim n \sigma_x \fx$, where $\sigma_x$ is the total
cross section, and $n$ is the gas density. Therefore, heat conduction
will dominates over distances no larger than $\Delta z \simlt F_{\rm
sat}/Q^+_x$, yielding the corresponding
maximum Thomson depth of
\begin{equation}
\Delta \tau_m = n \sigma_T \Delta z \sim {F_{\rm sat}\over Q^+_x} n
\sigma_T \simeq {P c \over\fx} {\sigma_T c_h \over \sigma_x c} \sim
10^{-3} T_8 ^{1/2} \;.
\label{tm}
\end{equation}
where $T_8$ is the temperature of the hot zone adjacent to the
temperature discontinuity in units of $10^8$ K (note that $\Delta z$
is essentially the ``Field length'' introduced by Begelman \& McKee
(1990) when the conductive flux is given by the saturated flux). Here
we made use of the fact that the transition from the hot to the cold
solution happens when $P c/\fx\sim$ few$^{-1}$.  Therefore, the direct
influence of heat conduction on the radiation transfer is negligible
because the Thomson depth of the region where conduction dominate is
quite small (similar conclusions were obtained by KMT in their
Appendix V). We in fact cannot afford to treat zones with such a small
optical depth for numerical reasons, and thus taking thermal
conduction into account is an expensive luxury for us.  On the other
hand, heat conduction may be quite important in determining what
solution the system picks in the unstable regime, but we believe that
we have taken its effects into account as discussed in \S
\ref{sect:multiplicity}.

\section{On the importance of  cloud formation}\label{sect:clouds}

In the case of a thermally unstable medium, the cold and hot phases
may actually be mixed, i.e., the cold mater may be in the form of
``clouds'' embedded in the hot surrounding medium. Physics of cloud
condensation and evaporation and radiative equilibria have been
previously studied by, e.g., Zel`dovich \& Pikel`ner 1969; Cowie \&
McKee 1977; KMT; Balbus 1985; Begelman \& McKee and McKee \& Begelman
(1990).  Although many aspects of the theory have been clarified in
these papers, much uncertainty remains due to the unknown distribution
of cloud sizes and separations. \rozanska \& Czerny (1996), \rozanska
(1999) studied the role of the thermal instability in X-ray
illuminated disks in AGN. They argued that instead of a sharp
transition from the hot to the cold solution, there exists a complex
two-phase medium with cold clouds embedded in a hotter gas.

However, the gas density of the cold phase is much larger (by a factor
of $\sim 100 - 10^3$) than that of the hot medium, so the cold clouds
experience a much greater gravitational force than the hot medium, and
thus they will sink, joining the cold (one-phase) medium below.  It
seems quite likely that the gravity may separate the two phases, such
that the hot phase will end up on the top of the cold phase just as a
light oil does if poured into water, thus leading to a sharp boundary
between the cold and the hot medium as suggested in our work here. In
addition, note that a cloud can also be broken up if its ``sinking
velocity'' is even a small fraction of the sound speed, since then the
face-on pressure is larger than the pressure on the sides of the
cloud.  In addition, the incident X-ray flux may be variable, and thus
can cause the height of the region where the two-phase medium may
exist to be constantly changing, leading to cloud destruction before
they can form. Finally, even if some clouds do form and are
efficiently maintained in a part of the illuminated layer, we believe
that their existence is marginal for the radiation transfer and the
resulting spectra. Indeed, in the case of the quasar emission line
regions, the clouds are thought to form above the accretion disk, in
the region filled with a hot Compton-heated gas (e.g., Krolik et
al. 1981). The latter, if optically thin, is basically invisible to
the observer.  Therefore, existence of any cool clouds embedded in
this hot matter is of a significant observational interest, since the
clouds are believed to be strong line emitters. In terms of a given
spectral line, then, a cloud looks like a bright emitting object in
the background of completely dark material.

The geometrical setup of the X-ray reprocessing problem is
substantially different. The cool clouds, if formed, exist in the
region {\em above} the cold thermally stable gas. Moreover, the
temperature of the clouds should be close to the temperature of the
cold stable phase below. This means that the emission spectrum of the
clouds will be close to that of the cold material below. Thus, in
terms of the observational signatures, except for increasing the
amount of the cool material somewhat, the clouds will be completely
washed out (unresolvable) on the background of the cold stable medium.
Accordingly, we believe that existence of the two-phase medium in a
region close to the temperature discontinuity is not only unclear, but
is of academic value at least as far as the X-ray reprocessing is
concerned. 

\section{Viscous dissipation and the hydrostatic equilibrium}

It has been pointed out to us by Julian Krolik that while the viscous
dissipation can be neglected in the energy equation of the
illuminated layer, one may still have to consider the effects of the
viscous heating for hydrostatic balance (equation \ref{heq1}).  This
is so because the gravity is canceled mainly by the radiation
pressure even in the illuminated layer of a radiation-dominated disk
(see \S \ref{sect:rd}). Thus, there could be a situation in which the
increase with height in the first term on the right in equations
(\ref{heq1}, \ref{heq2}) is entirely balanced by a corresponding
increase in the outward radiation pressure force (the third term),
which is exactly the case in a RD SS73 disk. One can show
that for this to be true, the gas density of the illuminated layer
must stay close to its midplane value (because viscous dissipation is
usually modeled as being proportional to the gas density -- see
SS73), a rather unlikely condition for the illuminated matter in the
setup of our paper. It is furthermore impossible if one resolves the
vertical disk structure more accurately. Simulations (Agol and Krolik
1999, private communications) show that convection changes the disk
profile in a way such that the gas density does not remain constant 
inside the RD disks; the gas density is greatest in the midplane and 
it decreases with height.

{}

\end{document}